\documentclass[12pt]{article}

\usepackage{amssymb,amsfonts,amsthm}
\usepackage{amsmath}
\usepackage{bm}             
\usepackage[mathscr]{eucal}
\usepackage{cancel}
\usepackage{wasysym}

\usepackage{graphicx}
\usepackage{subfigure}
\def\be{\begin{equation}}
\def\ee{\end{equation}}
\def\bea{\begin{eqnarray}}
\def\eea{\end{eqnarray}}

\def\bdelta{\mbox{\boldmath $\delta$}}

\def\hsp5{\hspace{5mm}}

\theoremstyle{remark}

\newcommand{\sfrac}[2]{{\textstyle{#1\over#2}}}

\title{\sc Dynamical systems in perturbative scalar field cosmology}

\begin{document}

\author{
\sc Artur Alho,$^{1}$\thanks{Electronic address:{\tt
artur.alho@tecnico.ulisboa.pt}}\,, Claes Uggla,$^{2}$\thanks{Electronic address:{\tt
claes.uggla@kau.se}}\,  and John Wainwright$^{3}$\thanks{Electronic
address:{\tt jwainwri@uwaterloo.ca}}\\
$^{1}${\small\em Center for Mathematical Analysis, Geometry and Dynamical Systems,}\\
{\small\em Instituto Superior T\'ecnico, Universidade de Lisboa,}\\
{\small\em Av. Rovisco Pais, 1049-001 Lisboa, Portugal.}\\
$^{2}${\small\em Department of Physics, Karlstad University,}\\
{\small\em S-65188 Karlstad, Sweden.}\\
$^{3}${\small\em Department of Applied Mathematics, University of Waterloo,}\\
{\small\em Waterloo, ON, N2L 3G1, Canada.}}


\date{}
\maketitle

\begin{abstract}

We derive a new \emph{regular} dynamical system on a 3-dimensional \emph{compact} state
space describing linear scalar perturbations of spatially flat Robertson-Walker geometries
for relativistic models with a minimally coupled scalar field with an exponential
potential. This enables us to construct the global solution space, illustrated
with figures, where known solutions are shown to reside on special
invariant sets. We also use our dynamical systems approach to obtain new results
about 
the comoving and uniform density curvature perturbations. Finally we show
how to extend our approach to more general scalar field potentials. This leads
to state spaces where the state space of the models with an exponential potential
appears as invariant boundary sets, thereby illustrating their role as building
blocks in a hierarchy of increasingly complex cosmological models.

\end{abstract}

\section{Introduction}

Inflation, quintessence --- scalar fields appear prominently in standard
cosmology, where the universe on large scales is described by a perturbed
Robertson-Walker (RW) background. But a lack of consensus, e.g. about
interpretation of observations, first principles, and issues such
as fine tuning of initial data, has resulted in a plethora of proposed
models and scalar field potentials, often connected with various heuristic
physical and mathematical considerations. This suggests that there might
be some value in a systematic approach using dynamical systems to analyze the
evolution of cosmological models on perturbed
RW backgrounds, especially those involving scalar fields.
With this goal in mind, we have embarked on a
research program using dynamical systems,
where this paper is the second in a series of papers (the first,
Alho \emph{et al} (2019)~\cite{alhetal19a}, treated scalar and
tensor perturbations of $\Lambda$CDM models).

A dynamical system consists of a system of autonomous nonlinear
first order ordinary differential equations (ODEs).
In the applications in cosmology that we have in mind
the state space ${\cal S}$ has a product structure
\begin{equation}
{\cal S}={\cal B}\times {\cal P},
\end{equation}
where ${\cal B}$ is the \emph{background state space}, which describes
the dynamics of a RW background, and ${\cal P}$ is the
\emph{perturbation state space}, which contains Fourier
decomposed gauge invariant variables that describe linear perturbations.
In this paper we, for brevity, neglect vector and tensor perturbation modes.
The system of differential equations has one subset of equations
that acts on ${\cal B}$ while the remaining equations
act on ${\cal P}$ with coefficients from ${\cal B}$. In this way the dynamics
in the background determine the dynamics of the perturbations.
A key step in our approach is to choose variables that lead to
\emph{regular equations on a bounded state space}. This
enables one to give a \emph{global}
description of the dynamics, in particular, the behaviour at early and
late times and the evolution at intermediate stages that may be
of physical interest. In addition the differential equations are well-suited
for performing systematic numerical simulations.

In Alho \emph{et al} (2019)~\cite{alhetal19a}
the background state space describes the $\Lambda$CDM model,
which represents the simplest situation since both ${\cal B}$ and ${\cal P}$
are one-dimensional sets, both for scalar and tensor perturbations.
In the present paper the background state space ${\cal B}$ describes
a spatially flat RW model with a minimally coupled scalar field. In order to show
how to incorporate such a source within our new framework
we have focused on the simplest potential, the exponential
potential, $V(\varphi)=V_0\exp(-\sqrt 6 \lambda\,\varphi),$
where $V_0$ and $\lambda$ are constants. In  this case
the background space ${\cal B}$  has dimension 2
and the perturbation space ${\cal P}$ has dimension 1, yielding a bounded state space
${\cal B}\times{\cal P}$ of dimension 3.
Using our approach we are able to give a complete picture of the dynamics,
showing all ways in which a model can evolve from early to late times,
illustrated by figures with representative orbits.
The value of the parameter $\lambda$ in the potential
determines two disparate families of models,
those with future acceleration and deceleration, respectively.

As regards earlier analytical work,
the perturbation equations have been solved explicitly in terms of Bessel functions,
by imposing the restriction that the scale factor $a$ has a power law
dependence on conformal time, which corresponds to a constant deceleration
parameter.\footnote{See for example Lyth and Stewart
(1992)~\cite{lytste92}, equations (21) and (22),
Durrer (2008)~\cite{dur08}, page 113,
and Weinberg (2008)~\cite{wei08} pages 480-482. These references describe
the primary application of the exponential scalar field potential in
inflationary cosmology, namely, determining the power spectrum for
perturbations in power law inflation.\label{fn:2} }
We show that these well known solutions
are described by orbits (solution trajectories)
on a two dimensional invariant subset of the three
dimensional state space ${\cal B}\times{\cal P}$ that contains
the future attractor of the whole state space.
These special solutions thus play a central role in the dynamics
of scalar fields with exponential potential, but do not describe
the full range of dynamic possibilities.

Our dynamical systems treatment also sheds light on the comoving
and uniform density curvature perturbations. In particular, it provides a new context to
recent discussions about the so-called ultra slow-roll inflation and conserved quantities,
see e.g.~\cite{rometal16,moopal15,tsawoo04,kin05,nametal13,cheetal13a,maretal13}.

The outline of the paper is as follows. In the next section we present the background
Einstein and (non-linear) Klein-Gordon (KG) equations for the spatially flat RW background with a
matter content given by a minimally coupled scalar field, and we give the
perturbed KG equation specialized to the uniform (flat) curvature gauge.
In Section~\ref{sec:dynsysscalarfield}
we formulate the background equations and the linearly perturbed KG equation as a
three-dimensional dynamical system with regular differential equations on a compact
state space ${\cal B}\times{\cal P}$.
This is followed by a dynamical systems analysis in
Section~\ref{sec:dynsysanalysexp}, where figures with representative orbits globally
illustrate the entire state space.
In addition we present analytic asymptotic descriptions of the orbits
at late and early times.
In Section~\ref{bessel} we review the well known explicit solutions
and relate them to our dynamical systems framework.
In Section~\ref{sec:cons} we derive new analytic asymptotic results for
the comoving and uniform density curvature perturbations
and relate them to recent research
in the case of ultra slow-roll inflation.
Section~\ref{sec:disc} shows how our dynamical systems
approach can be extended to a hierarchy of increasingly complex models.

\section{Field equations for a minimally coupled scalar field}

\subsection{The background equations  \label{background}}

We consider a spatially flat RW background with a metric written as
\begin{equation}
ds^2 = - dt^2 + a^2\gamma_{ij} dx^i dx^j = a^2\left(- d\eta^2 + \gamma_{ij} dx^i dx^j \right)
= - H^{-2}dN^2 + a^2\gamma_{ij} dx^i dx^j,
\end{equation}
where $a$ is the background scale factor, $H$ the Hubble variable,
and $\gamma_{ij}$ is the flat spatial 3-metric, which in Cartesian
coordinates is given by $\delta_{ij}$. The different time coordinates
above are the clock time $t$, the conformal time $\eta$, and the $e$-fold time
\begin{equation}
N = \ln(a/a_0),
\end{equation}
where $N$ describes the number of background $e$-foldings
with respect to some reference epoch at which $a=a_0$ and hence $N=0$.

The background equations that govern the matter distribution in
a flat RW universe are the Friedman equation and the energy
conservation equation
\begin{equation} \label{matter}
\rho=3H^2, \qquad {\dot\rho} =-3H (\rho+p).
\end{equation}
where overdot denotes derivative with respect to $t$.

The energy density $\rho$ and pressure $p$ for a minimally
coupled scalar field $\varphi$ with a
non-negative potential, $V = V(\varphi)$ are given by\footnote{See for
example Liddle and Lyth (2000)~\cite{lidlyt00}, equations (3.3) and (3.4).}
\begin{equation} \label{scalar.field}
\rho=\frac12{\dot\varphi}^2 +V(\varphi), \qquad
\rho + p={\dot\varphi}^2,
\end{equation}
Making the obvious substitutions of equations~\eqref{scalar.field}
into~\eqref{matter} we obtain\footnote{See for example Liddle and
Lyth (2000)~\cite{lidlyt00}, equations (3.5) and (3.6).}
\begin{equation} \label{gov.t}
\ddot{\varphi} + 3H\dot{\varphi} + V_{,\varphi}=0, \qquad
3H^2 = \frac12\dot{\varphi}^2 + V(\varphi),
\end{equation}
where $_{,\varphi}$ denotes $\frac{d}{d\varphi}$.
Given a potential $V(\varphi)$ these equations
determine $\varphi(t)$ and $H(t)$ when $\varphi$ and ${\dot\varphi}$
are specified at an initial time.
The equation $\dot{a} = aH$ then determines the scale factor $a$ as a quadrature.

Before continuing we digress to introduce the deceleration parameter $q$ which
is useful in describing the dynamics:
\begin{equation} \label{def.q}
{\dot H}=-(1+q){H^2}.
\end{equation}
It follows from~\eqref{matter} and~\eqref{scalar.field} that
\begin{equation} \label{q.varphi}
1+q = \frac12\frac{{\dot \varphi}^2}{H^2}.
\end{equation}

We find it convenient to use the $e$-fold time $N$
instead of the clock time $t$ as our starting point and to introduce the following
quantities:\footnote{The factors of $\sqrt 6$ are included in the definitions~\eqref{defback1}
in order to simplify future algebra. The reason for using the
notation $\Sigma$ for the kernel is because this variable plays a similar role as Hubble-normalized shear,
which is typically denoted with the kernel $\Sigma$, see e.g.~\cite{waiell97}.}
\begin{equation}\label{defback1}
\Sigma_\varphi = \frac{1}{\sqrt{6}}\varphi', \qquad
\Omega_V = \frac{V}{3H^2}, \qquad \lambda = -\frac{1}{\sqrt{6}}\frac{V_{,\varphi}}{V},
\end{equation}
where a ${}^\prime$ denotes the $e$-fold time derivative
$\frac{d}{dN} = H^{-1}\frac{d}{dt}$.
Equations~\eqref{def.q} and~\eqref{q.varphi} assume the form
\begin{equation} \label{q.phi}
H' =-(1+q)H, \qquad 1+q = 3\Sigma_\varphi^2.
\end{equation}
The definitions~\eqref{defback1} have the effect of eliminating the explicit appearance of
$H$ in the governing equations~\eqref{gov.t}. The second equation
becomes
\begin{equation}
 \Sigma_\varphi^2 + \Omega_V =1,\label{Gauss}
\end{equation}
while the second order ODE is replaced by two coupled first order ODEs:\footnote{As
a first step divide both equations by $H^2$ and note that
\begin{equation*}
\frac{d}{dt}=H\frac{d}{dN},\quad \frac{d^2}{dt^2}=
H^2\left(\frac{d^2}{dN^2} -(1+q)\frac{d}{dN}\right).
\end{equation*}
Use~\eqref{q.phi}  to eliminate $1+q$, and
use  the definition of $\lambda$ in~\eqref{defback1} and~\eqref{Gauss}
to eliminate $V_{,\varphi}$
in~\eqref{gov.t}.}
\begin{subequations}\label{backgroundeq}
\begin{align}
\varphi^\prime &= \sqrt{6}\Sigma_\varphi ,\label{phisig}\\
\Sigma_\varphi^\prime &= 3(1 - \Sigma_\varphi^2)(\lambda - \Sigma_\varphi).\label{Sigmaeq}
\end{align}
\end{subequations}

In general the scalar $\lambda $,  defined by~\eqref{defback1}, is a function
of $\varphi$
but in the case of an exponential potential
$\lambda$ is constant. We use this constant value to  define $V$
according to\footnote{This labelling of $V$ is consistent with the
definition~\eqref{defback1} of $\lambda$.}
\begin{equation}\label{Vpot}
V = V_0\exp(-\sqrt{6}\lambda\varphi), \qquad V_0> 0.
\end{equation}
Since $\lambda$ is now a constant~\eqref{phisig}
decouples from~\eqref{Sigmaeq}, and can be used to determine
$\varphi$ by quadrature once~\eqref{Sigmaeq} has been solved.
Thus~\eqref{Sigmaeq} contains the essential dynamical information for the background scalar field in the case of an exponential potential.
Since $V$ has been defined to be positive, it follows from~\eqref{defback1}
that $\Omega_V$ is positive and hence
from~\eqref{Gauss}  that $\Sigma_\varphi$ \emph{is bounded}:
$-1 < \Sigma_\varphi <1$.  However,
since the right hand side of~\eqref{Sigmaeq}  is continuously differentiable and
zero when $\Sigma_\varphi=\pm1$
it is possible and desirable to extend the range of $\Sigma_\varphi$ to
include the invariant boundary values $\Sigma_\varphi = \pm 1$:
\begin{equation}
-1\leq\Sigma_{\varphi}\leq 1.
\end{equation}
%
Equation~\eqref{q.phi} now restricts the deceleration parameter:
\begin{equation}
-1 \leq q \leq 2,
\end{equation}
where $q=-1$ if $\Sigma_\varphi = 0$ and $q=2$ if $\Sigma_\varphi = \pm 1$.

%
%

\subsection{The perturbed Klein-Gordon equation}

To describe the linear perturbations of the scalar field we use the perturbed
KG equation, which has a particularly simple form if one
uses as dependent variable the linear perturbation of the
scalar field in the uniform (flat) curvature gauge,\footnote{Once one 
chooses the uniform curvature gauge or the comoving gauge
as we do in section~\ref{sec:cons},
there is no remaining gauge freedom. See for example~\cite{malwan09}
or~\cite{uggwai19a} section 3.}
which we denote by $\varphi_{\mathrm c}$. An advantage of using the
uniform curvature gauge is that the metric perturbations can be eliminated
from the KG equation using the Einstein equations, leaving a closed equation
which can be written as follows (see, e.g., Uggla and Wainwright
(2019)~\cite{uggwai19c}, section 3, Appendix A.3 and in particular
equation (A.21)):
\begin{equation} \label{KG.general}
\partial_N^2{\varphi}_{\mathrm c} +
\frac{V}{H^2} \partial_N {\varphi}_{\mathrm c} +
\frac{(V_{,\varphi\varphi}+2\varphi' V_{,\varphi} +
(\varphi' )^2V)}{H^2}{\varphi}_{\mathrm c} -
{\cal H}^{-2}\,{\bf D}^2{\varphi}_{\mathrm c}=0,
\end{equation}
where $\varphi$ is the background scalar field,
\begin{equation} \label{cal H}
{\cal H} = a H,
\end{equation}
and ${\bf D}^2$ is the background spatial Laplacian associated with $\gamma_{ij}$.  For an
exponential potential~\eqref{Vpot} the KG equation assumes the form
\begin{equation}\label{KG2scalarsexp}
\partial_N^2\varphi_\mathrm{c} + 3(1 -  \Sigma_\varphi^2)\partial_N\varphi_\mathrm{c}
 + 18(1 - \Sigma_\varphi^2)(\lambda - \Sigma_\varphi)^2\varphi_\mathrm{c}
- {\cal H}^{-2}{\bf D}^2\varphi_\mathrm{c}= 0 ,
\end{equation}
in the notation of~\eqref{defback1}.

\section{Derivation of the dynamical system\label{sec:dynsysscalarfield}}
To obtain a dynamical system, i.e., an autonomous closed system of first order ODEs,
that describes the evolution of the perturbations,
we first introduce Cartesian spatial coordinates and make
a spatial Fourier decomposition of the perturbation variables. This results in
\begin{equation}\label{Dk}
{\cal H}^{-2}{\bf D}^2 \rightarrow - k^2{\cal H}^{-2},
\end{equation}
thereby transforming~\eqref{KG2scalarsexp}
to
\begin{equation}\label{FourierKG2scalarsexp}
\varphi_\mathrm{c}'' + 3(1 -  \Sigma_\varphi^2)\varphi_\mathrm{c}'
+ 18(1 - \Sigma_\varphi^2)(\lambda - \Sigma_\varphi)^2\varphi_\mathrm{c}
+ k^2{\cal H}^{-2}\varphi_\mathrm{c}= 0 ,
\end{equation}
where $k$ is the wave number.
The Fourier coefficients of $\varphi_\mathrm{c}$ are labelled by an
index $k$, $(\varphi_\mathrm{c})_k$, but for brevity we will drop the
index $k$ when there is no danger of confusion.
At this stage we have to take into account that the
Fourier coefficients $\varphi_\mathrm{c}$
are \emph{complex} functions.\footnote{The physical perturbation
$\varphi_\mathrm{c}$ of the scalar field is assumed
to be real. Note also that the Fourier coefficients $\varphi_{\mathrm{c},{\bf k}}$ and
$\varphi_{\mathrm{c},{-\bf k}}$, where ${\bf k}$ is the wave vector, obey the same equation
since the wave vector ${\bf k}$ only enters the problem via the wave number as $k^2$, which
is why we have denoted the Fourier coefficients by $k$. For a discussion on
Fourier decompositions for cosmological perturbations,
see~\cite{alhetal19a} and references therein.}
Since we want to obtain a dynamical system
in terms of real variables we write
$\varphi_\mathrm{c}=f_1+if_2$, where $f_1$ and $f_2$ are
\emph{real} functions, labelled by the wave number $k$,
which describe the general solution for a given $k$. 
Since~\eqref{FourierKG2scalarsexp}
is a linear differential equation with \emph{real} coefficients it follows that
$f_1,f_2$ satisfy this differential equation which we rewrite as
\begin{equation}\label{FourierKG.real}
f'' + 3(1 -  \Sigma_\varphi^2)f'
+ 18(1 - \Sigma_\varphi^2)(\lambda - \Sigma_\varphi)^2 f
+ k^2{\cal H}^{-2} f= 0,
\end{equation}
where $f$ stands for $f_1$ or $f_2$.  We now consider
$f$ and $f'$, with subscript $1$ or $2$, as independent
variables and represent them using polar coordinates
$f_i = r_i\cos\theta_i$, $f'_i = r_i\sin\theta_i$, for $i=1,2$,
where $\theta$ has period $2\pi$.
Note that the variables $f$ and $\theta$ are related by the equation
\begin{equation}\label{f.theta}
f' = f \tan\theta,
\end{equation}
for both values of the index $i$.
Differentiating this equation to obtain $f''$
converts equation~\eqref{FourierKG.real} into the following
first order differential equation for $\theta$:
\begin{equation}\label{thetavarphi}
\theta^\prime =  - \sin^2\theta
-  3(1 - \Sigma_\varphi^2)\sin\theta\cos\theta
-\left[18(1 - \Sigma_\varphi^2)(\lambda - \Sigma_\varphi)^2 + k^2{\cal H}^{-2}\right]\cos^2\theta,
\end{equation}
where $\theta$ stands for $\theta_1$ or $\theta_2$.
The angular variable $\theta$ will be useful for describing the global
structure of the state space.
For doing local calculations, however, it is more convenient to use
the variable $y$ defined by
\begin{equation}\label{theta.y}
y = \frac{f'}{f} = \tan\theta,
\end{equation}
where $y$ stands for $y_1$ or $y_2$. Since $y'=(\sec^2\theta) \theta',$
equation~\eqref{thetavarphi} leads to the Riccati equation
\begin{equation}\label{yphi}
y^\prime = -\left[y^2 +  3(1 - \Sigma_\varphi^2)y +
18(1 - \Sigma_\varphi^2)(\lambda - \Sigma_\varphi)^2\right]  - k^2{\cal H}^{-2}.
\end{equation}
In summary, a \emph{complex solution} $\varphi_{\mathrm c}(N)=f_1(N)+if_2(N)$ of
the KG equation~\eqref{FourierKG2scalarsexp}, which is a second order linear
differential equation, is described by
\emph{two real solutions} $y_1(N), y_2(N)$ of the first order
non-linear differential equation~\eqref{yphi}. The functions $y_i$ determine the
functions $f_i$ by quadrature using~\eqref{theta.y}, which yields
\begin{equation} \label{y.det.f}
f_i(N)=f_i(0) \exp\left[\int_0^N y_i(\tilde N)d{\tilde N}\right],
\end{equation}
where the initial values $f_i(0)$ are determined by the initial value of
$\varphi_{\mathrm c}$ at $N=0$.

Equations~\eqref{thetavarphi} and~\eqref{yphi} are first order
non-autonomous ODEs with time dependent coefficients that depend on the
background solutions for $\Sigma_\varphi$ and ${\cal H}^{-2}$. The
wave number $k$ only appears in the equations as a parameter
in the dimensionless combination $k^2{\cal H}^{-2}$. To obtain a
dynamical system, i.e., an autonomous closed system of first order ODEs,
we need to augment the ODE~\eqref{yphi}
for $y$ (or~\eqref{thetavarphi} for $\theta$) by including the
ODE~\eqref{Sigmaeq} for $\Sigma_{\varphi}$ and an ODE for $k^2{\cal H}^{-2}$
when $k \neq 0$. We therefore introduce a new variable $Z$ (referred to as the physical
wave number in Hubble units in~\cite{tsawoo04}):
\begin{equation}\label{calH2}
Z = k^2{\cal H}^{-2},
\end{equation}
which satisfies the differential equation
\begin{equation}\label{calH2diff}
Z^\prime = 2qZ = 2(3\Sigma_\varphi^2 - 1)Z,
\end{equation}
on account of~\eqref{q.phi}, \eqref{cal H} and $a^\prime=a$.

For the state space described by $(\Sigma_\varphi, Z, y)$
we thereby obtain the following dynamical system:
\begin{subequations}\label{dynsystotN}
\begin{align}
\Sigma_\varphi^\prime &= 3(1 - \Sigma_\varphi^2)(\lambda - \Sigma_\varphi),\label{Sigmaeq4}\\
Z^\prime &= 2(3\Sigma_\varphi^2 - 1)Z,  \label{Zeq4}\\
y^\prime &= -\left[y^2 +  3(1 - \Sigma_\varphi^2)y +
18(1 - \Sigma_\varphi^2)(\lambda - \Sigma_\varphi)^2\right]  - Z,
\end{align}
\end{subequations}
while for the state space described by $(\Sigma_\varphi, Z, \theta)$, where we recall that
$y=\tan\theta$, the above equation for $y$ is replaced with
\begin{equation}\label{thetaeq}
\theta^\prime = -\left[\sin^2\theta +  3(1 - \Sigma_\varphi^2)\sin\theta\cos\theta +
18(1 - \Sigma_\varphi^2)(\lambda - \Sigma_\varphi)^2\cos^2\theta\right]  - Z\cos^2\theta .
\end{equation}
The dynamical systems for the state spaces $(\Sigma_\varphi, Z, y)$ and $(\Sigma_\varphi, Z, \theta)$
are regular in the sense that the expressions on the right hand sides
are differentiable functions of the state space variables. Note also that the two
ODEs~\eqref{Sigmaeq4} and~\eqref{Zeq4}, which describe the background
state space $(\Sigma_\varphi, Z)$, are uncoupled since $Z$ does not
appear in~\eqref{Sigmaeq4}. Thus one can solve~\eqref{Sigmaeq4} for
$ \Sigma_\varphi$ and then express $Z$ as a quadrature using~\eqref{Zeq4}.

Before continuing we comment on the interpretation of the variable $Z = k^2{\cal H}^{-2}$.
Perturbations that satisfy $k^2{\cal H}^{-2}\ll1$ are called long wavelength
or super-horizon, while those that satisfy $k^2{\cal H}^{-2}\gg1$ are said to be
short wavelength. Long wavelength perturbations are usually studied by
choosing the idealized limiting value $k=0$,\footnote{The long wavelength
case ($k=0$) can be solved
explicitly for \emph{any} scalar field potential, even up to second order, as shown
in~\cite{uggwai19c}.} which corresponds to $Z=0$.
On the other hand, short wavelength perturbations
correspond to $Z\rightarrow \infty$. We also note that in choosing
$Z=k^2{\cal H}^{-2}$ as a dynamical variable we have in a
sense `hidden' the wave number $k$ when formulating the dynamical system. However,
if we choose the reference time $t_0$ (i.e., when $N=0$) to be the time for setting
initial data in the state space $(\Sigma_\varphi,Z, y)$ (or $(\Sigma_\varphi,Z, \theta)$),
then different choices of $Z_0 = k^2{\cal H}_0^{-2}$ for a given ${\cal H}_0$
yield solutions with different wave number $k$.

A drawback of the system~\eqref{dynsystotN} is that the
variable $Z$ can become unbounded for some models.\footnote{The variable $Z$ becomes unbounded
when $k\neq 0$ and ${\cal H} \rightarrow 0$. However,
since $q$ is bounded for the present models ${\cal H} \rightarrow 0$ only asymptotically
toward the past or the future, depending on the asymptotic
signs of $q$.} To deal with this
situation we introduce the bounded variable
\begin{equation} \label{bar.Z}
\bar{Z} = \frac{Z}{1 + Z} = \frac{k^2}{k^2 + {\cal H}^2}, \qquad Z = \frac{\bar{Z}}{1 - \bar{Z}},
\end{equation}
which due to~\eqref{calH2diff} obeys the equation
\begin{equation}
\bar{Z}^\prime = 2(3\Sigma_\varphi^2 - 1)\bar{Z}(1 - \bar{Z}),
\end{equation}
where the long (short) wavelength limit corresponds to $\bar{Z}=0$ ($\bar{Z}=1$).
This change of variable, however, leads to the appearance of the term
$\bar{Z}/(1 - \bar{Z})$, which is unbounded
when $\bar{Z}=1$, in the equations for $y$ and $\theta$.
To regularize the dynamical system we choose a new time variable $\bar{N}$ according to
\begin{equation}
\frac{d\bar{N}}{dN} = \frac{1}{1 - \bar{Z}} = 1 + Z.
\end{equation}
This results in the following regular dynamical system:
\begin{subequations}\label{gensysscalar}
\begin{align}
\frac{d\Sigma_\varphi}{d\bar{N}} &= 3(1 - \Sigma_\varphi^2)(\lambda - \Sigma_\varphi)(1 - \bar{Z}), \label{gen1}\\
\frac{d\bar{Z}}{d\bar{N}} &= 2(3\Sigma_\varphi^2 - 1)\bar{Z}(1 - \bar{Z})^2,\label{gen2} \\
\frac{dy}{d\bar{N}} &= -\left[y^2 +  3(1 - \Sigma_\varphi^2)y +
18(1 - \Sigma_\varphi^2)(\lambda - \Sigma_\varphi)^2\right](1 - \bar{Z})  - \bar{Z},
\label{gen3}
\end{align}
\end{subequations}
on the state space $(\Sigma_{\varphi}, {\bar Z}, y)$ defined by
\begin{equation}\label{state.space.unb}
-1\leq \Sigma_{\varphi}\leq1, \qquad 0\leq{\bar Z}\leq 1, \qquad
-\infty < y < \infty.
\end{equation}
If we replace $y$ by the angular variable $\theta$
of period $2\pi$, equation~\eqref{gen3} is replaced by
\begin{equation}\label{thetaeq2}
\begin{split}
\frac{d\theta}{d\bar{N}} &= -\left[\sin^2\theta +  3(1 - \Sigma_\varphi^2)\sin\theta\cos\theta +
18(1 - \Sigma_\varphi^2)(\lambda - \Sigma_\varphi)^2\cos^2\theta\right](1 - \bar{Z})\\  & \quad - \bar{Z}\cos^2\theta,
\end{split}
\end{equation}
which results in a state space that is bounded (and compact).\footnote{At this stage the reader
might ask why we used the variable $\varphi_\mathrm{c}$ and the KG equation as our starting point
and not metric perturbation variables and the Einstein field equations in, e.g., the uniform curvature
gauge. The reason is that in the latter case $\Sigma_\varphi$ occurs in the denominator,
which leads to that the perturbed Einstein field equations break down when $\Sigma_\varphi =0$.}

\newpage
\section{Analysis of the dynamical system}\label{sec:dynsysanalysexp}

In this section we use the dynamical system that is described
by the differential equations~\eqref{gensysscalar} and~\eqref{thetaeq2}
on the state space ${\cal S}={\cal B}\times{\cal P}$ to give a
complete description of the dynamics of a perturbed scalar field
with exponential potential.
The dynamical system depends on the parameter $\lambda$ which we
assume satisfies $0 \leq \lambda<1$, since these are the values of
primary physical interest.\footnote{Firstly, there is no restriction in
assuming that $\lambda$ is non-negative,
since the field equations are invariant under the transformation
$(\varphi,\Sigma_\varphi) \rightarrow -(\varphi,\Sigma_\varphi)$ and
$\lambda \rightarrow -\lambda$. Secondly,
the fixed point $\Sigma_{\varphi}= \lambda$ is the future attractor for
the $\Sigma_{\varphi}$ background state space. At this fixed point the deceleration
parameter $q=3\lambda^2-1$ and is thus constant. The range $0\leq\lambda\leq 1$
corresponds to the range $-1\leq q\leq 1$ for $q$. This range of $q$ also
describes a space-time with a perfect fluid with a linear equation of state $p=w\rho$,
with $w$ in the range $-1\leq w\leq 1$.
Thus $\lambda=0$ corresponds to a cosmological constant while the bifurcation
value $\lambda=1$ corresponds to a stiff fluid with speed of
sound equal to that of light. On the other hand $\lambda>1$ yields an equation
of state with superluminal speed.}

\subsection{Invariant sets and fixed points in ${\cal B}\times{\cal P}$}

The state space ${\cal S}={\cal B}\times{\cal P}$ is a three-dimensional product
space where ${\cal B}$ is the background state space, with coordinates
$\Sigma_{\varphi}$ and ${\bar Z}$, subject to
\begin{equation}\label{bg.rectangle}
-1\leq \Sigma_{\varphi}\leq 1, \qquad
0\leq {\bar Z}\leq 1,
\end{equation}
while ${\cal P}$ is the perturbation state
space, which is a circle parameterized by
the angular coordinate $\theta$ of period $2\pi$.
We can visualize ${\cal S}$
by rotating the rectangle~\eqref{bg.rectangle} through $2\pi$ radians about a line
$ \Sigma_{\varphi}=b<-1$ to form a solid finite cylinder with a cylindrical hole
along its axis. The surfaces
of constant $\Sigma_{\varphi}$ are coaxial cylinders whose common axis defines
the $\bar{Z}$-axis with $0\leq\bar{Z}\leq1$.
The cylinders $ \Sigma_{\varphi}=1$ and $\Sigma_{\varphi}=-1$
form the outer and inner boundaries of the solid cylinder, respectively,
and are invariant sets of the dynamical system,
as follows from equation~\eqref{gen1}. The planes
${\bar Z}=0$ and ${\bar Z}=1$ form the bottom and top of the solid cylinder,
respectively, and are also invariant sets as follows from equation~\eqref{gen2}.
In this representation  \emph{the state space ${\cal B}\times{\cal P}$ is a
compact subset of ${\mathbb R}^3$}.
On the other hand if we use the perturbation variable $y=\tan\theta$ then
the state space ${\cal B}\times{\cal P}$ is
the infinite slab in ${\mathbb R}^3$ defined by the
inequalities~\eqref{state.space.unb}.\footnote{Since
$\tan (\theta +\pi)=\tan \theta$, the mapping
$y = \tan\theta$ is two-to-one and therefore
when $\theta$ makes one revolution ($0\rightarrow 2\pi$)
$y$ has to be traversed twice $-\infty \rightarrow +\infty$.}

The invariant sets $\Sigma_{\varphi}=1$ and
$\Sigma_{\varphi}=-1$ describe the limiting case of a massless scalar field
model ($V=0$, see equations~\eqref{defback1}
and~\eqref{backgroundeq}). We will therefore refer to the invariant sets
$\Sigma_{\varphi}=\pm1$ as the two components of the
\emph{massless scalar field boundary}.
Because of the physical interpretation
of ${\bar Z}$ (see section~\ref{sec:dynsysscalarfield}) we
refer to the invariant set ${\bar Z}=0$
as the \emph{long wavelength boundary},
and ${\bar Z}=1$ as the \emph{short wavelength boundary}.
In addition to the above invariant boundary sets there is an important
invariant set given by the interior cylinder with $\Sigma_{\varphi}=\lambda$.

We obtain the fixed points of the dynamical system by equating the
right hand sides of~\eqref{gensysscalar} to zero. It follows
that if $0 \leq \lambda < 1$ and $\lambda \neq1/\sqrt{3}$ then all the fixed points of the
dynamical system lie in the long and short wavelength boundary sets
$\bar{Z} = 0$ and $\bar{Z}=1$. On the $\bar{Z}=0$ boundary they either satisfy
$\Sigma_\varphi = \lambda$ or $\Sigma_\varphi^2=1$.
There are two hyperbolic fixed points with
$\Sigma_{\varphi} = \lambda$ which we
denote by $\mathrm{A}_{\lambda}$ and $\mathrm{S}_{\lambda}$:
\begin{subequations}  \label{fixed.points.A,S}
\begin{align}
\mathrm{A}_{\lambda}\!:\qquad\Sigma_{\varphi}&=\lambda,\quad \bar{Z}=0, \quad y=0,\\
\mathrm{S}_{\lambda}\!:\qquad \Sigma_{\varphi}&=\lambda,\quad \bar{Z}=0,
\quad y=-3(1-\lambda^2).
\end{align}
\end{subequations}
There are also two non-hyperbolic fixed points
denoted by $\mathrm{M}_{\pm}$, given by
\begin{equation} \label{fixed.points.M}
\mathrm{M}_{\pm}\!: \qquad \Sigma_{\varphi} = \pm 1, \quad \bar{Z}=0, \quad y=0.
\end{equation}
However, when we draw pictures of the global state space we will use $\theta$ as
the perturbation variable, with $\tan \theta=y$. Then $y=0$ corresponds
to $\theta=n\pi, n=0,1$, where $y=-3(1-\lambda^2)$ corresponds
to $\theta=\chi + n\pi, n=0,1$,
where $\chi=-\arctan 3(1-\lambda^2).$ In other words, each of the fixed
points~\eqref{fixed.points.A,S} and~\eqref{fixed.points.M} appear
twice as $\theta$ makes one revolution ($0\rightarrow 2\pi$).

The short wavelength boundary ${\bar Z}=1$ also plays an important role.
On ${\bar Z}=1,$ using $\theta$ as the perturbation variable, the differential
equations~\eqref{gensysscalar} and~\eqref{thetaeq2} simplify to
\begin{equation}\label{thetabareq2}
\frac{d\Sigma_{\varphi}}{d\bar{N}} = 0,
\qquad  \frac{d\theta}{d\bar{N}} = -\cos^2\theta .
\end{equation}
There are therefore two lines of (non-hyperbolic) fixed points at ${\bar Z}=1$ with
constant $\Sigma_\varphi$ given by $\cos\theta = 0 $, i.e., $\theta = \pi/2 + n\pi, n=0,1$.
Except at these fixed points, $\theta$ is monotonically decreasing on ${\bar Z}=1$
at constant $\Sigma_\varphi$. It follows that the invariant set ${\bar Z}=1$ is spanned
by a family of heteroclinic cycles (concentric circles) given by
$\Sigma_{\varphi} = \mathrm{const}.$, with
$-1\leq \Sigma_{\varphi}\leq 1$. At $\theta = \pi/2 + n\pi$
equation~\eqref{thetaeq2} takes the form
\begin{equation}\label{thetaeq2fixed}
\left.\frac{d\theta}{d\bar{N}}\right|_{\cos\theta = 0} = -(1 - \bar{Z}),
\end{equation}
which implies that orbits in the vicinity of $\bar{Z}=1$ shadow
the heteroclinic cycles and do
not end at the fixed points on these cycles. Moreover,
as we will see, asymptotics are associated with
$\Sigma_\varphi = \lambda, \pm 1$, and hence it is only the heteroclinic
cycles at these values of
$\Sigma_\varphi$ which are of asymptotic relevance.
We will denote these heteroclinic cycles
by ${\cal H}_\lambda$, ${\cal H}_{\pm1}$, respectively
(this use of $\cal H$ with subscripts should not be confused with ${\cal H} = aH$).

\begin{figure}[ht!]
	\begin{center}
			\includegraphics[width=0.45\textwidth]{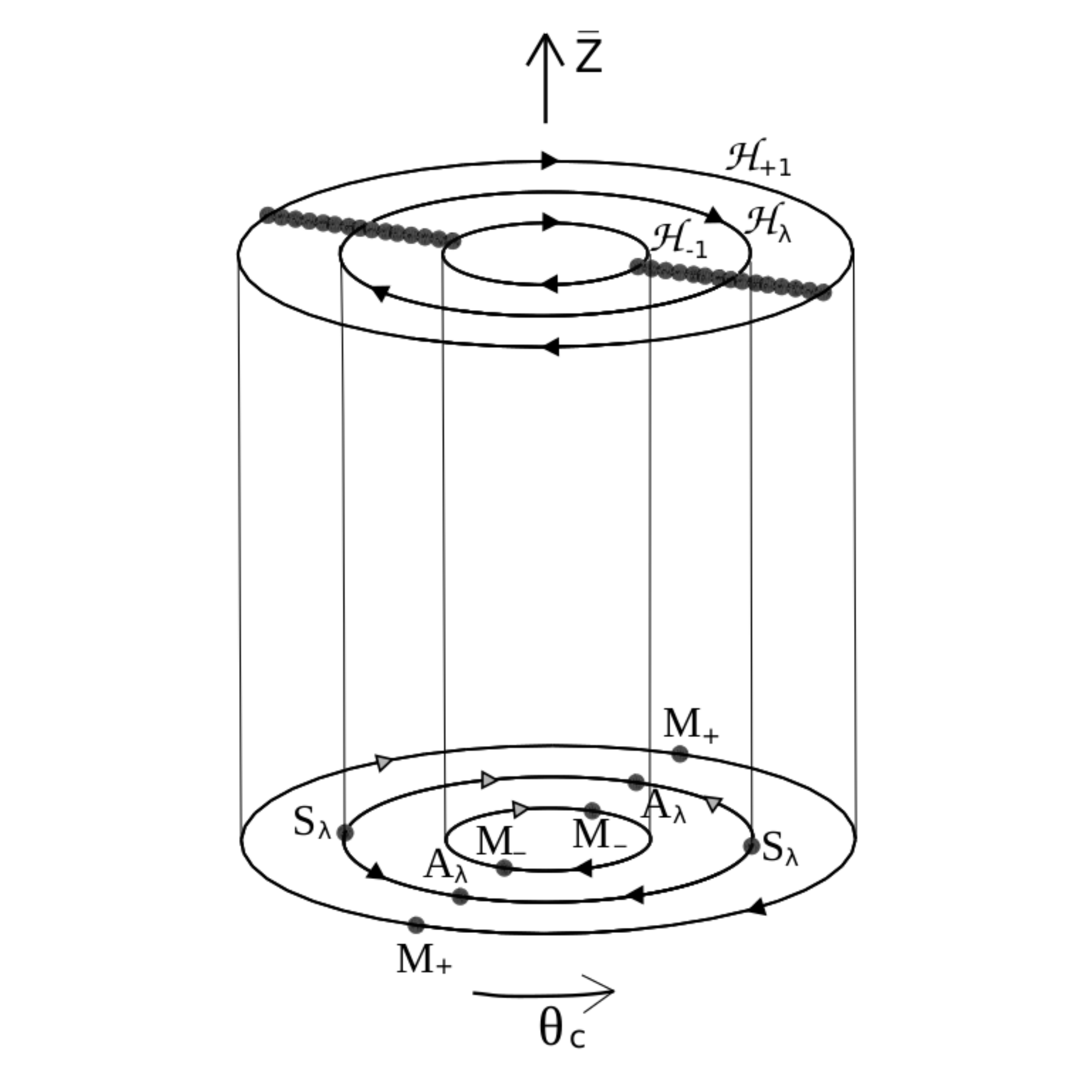}
		\vspace{-1cm}
	\end{center}
		
	\caption{Invariant sets and fixed points in the compact state space
	${\cal S}={\cal B}\times{\cal P}$.
             The three coaxial cylinders
             whose common axis is the ${\bar Z}-$ axis with $0\leq{\bar Z}\leq 1$
             are the invariant cylinders $\Sigma_\varphi=-1,\lambda, 1$.
             The isolated fixed points
             $\mathrm{A}_{\lambda}, \mathrm{S}_{\lambda}$ and $\mathrm{M}_{\pm1}$
             lie in the long wavelength boundary $\bar{Z}=0$
             and the short wavelength boundary $\bar{Z}=1$
             contains the heteroclinic cycles
             ${\cal H}_{-1}$, ${\cal H}_{\lambda}$, ${\cal H}_{1}$.}
\label{Figstatespace}
\end{figure}

In Fig.~\ref{Figstatespace} we show the key invariant sets
and fixed points in the state space ${\cal S}={\cal B}\times{\cal P}$.
%

\subsection{Orbit structure of the background state space ${\cal B}$\label{background}}

The background state space $\cal B$ with coordinates $(\Sigma_\varphi, {\bar Z})$
 is defined by the inequalities
\begin{equation}\label{BoundRect}
-1\leq \Sigma_{\varphi}\leq 1, \qquad
0\leq {\bar Z}\leq 1.
\end{equation}
The differential equations that define a dynamical
system on ${\cal B}$ are the first two equations of
the system~\eqref{gensysscalar}. In the background it is convenient to
use the usual $e$-fold time $N$ instead of the modified time
${\bar N}$, leading to the following equations:
\begin{subequations}\label{backgroundsys2}
\begin{align}
\Sigma_\varphi^\prime &= 3(1 - \Sigma_\varphi^2)(\lambda - \Sigma_\varphi),\label{Sigmaeq4.2}\\
\bar{Z}^\prime &= 2(3\Sigma_\varphi^2 - 1)\bar{Z}(1-\bar{Z}).\label{barZN}
\end{align}
\end{subequations}
The fixed points of this dynamical system can be found by inspection.
If $\lambda\neq 1/\sqrt 3$ there are six isolated fixed points,
which lie on the boundary of the rectangle~\eqref{BoundRect}:
\begin{subequations}
\begin{align}
{\bar Z}=&0,  \qquad \Sigma_\varphi=\pm 1, \lambda,\\
{\bar Z}=&1,  \qquad \Sigma_\varphi=\pm 1, \lambda .
\end{align}
\end{subequations}
The fixed points ${\bar Z}=0,\, \Sigma_\varphi=\pm 1$
are local sources and form the past attractor, while the
fixed points ${\bar Z}=1,\Sigma_\varphi=\pm 1$
are saddles for all values of $\lambda$.

The future attractor  depends on the value of the
parameter $\lambda$, and is given by
the following two fixed points:
\begin{subequations}
\begin{alignat}{10}
\text{i)}\quad {\bar Z} &=& 0,  \qquad \Sigma_\varphi &=& \lambda, \quad
&\text{a sink if}& \quad 0 &\leq & \lambda &<& \frac{1}{\sqrt{3}},\\
\text{ii)}\quad {\bar Z} &=& 1,  \qquad \Sigma_\varphi &=& \lambda, \quad
&\text{a sink if}& \quad \frac{1}{\sqrt{3}} &<& \lambda &<& 1.\,\,\,\,\,
\end{alignat}
\end{subequations}
The line $\Sigma_\varphi = \lambda\neq 1/\sqrt{3}$ is a special orbit
joining the fixed points i) and ii).
If $\lambda = 1/\sqrt{3}$ (and therefore $q=0$) this line becomes
a line of fixed points, which corresponds to a bifurcation that
transfers stability from i) to ii) as $\lambda$ increases.

\begin{figure}[ht!]
	\begin{center}
		\subfigure[$\lambda=1/\sqrt{6}$]{\label{}
			\includegraphics[width=0.51\textwidth]{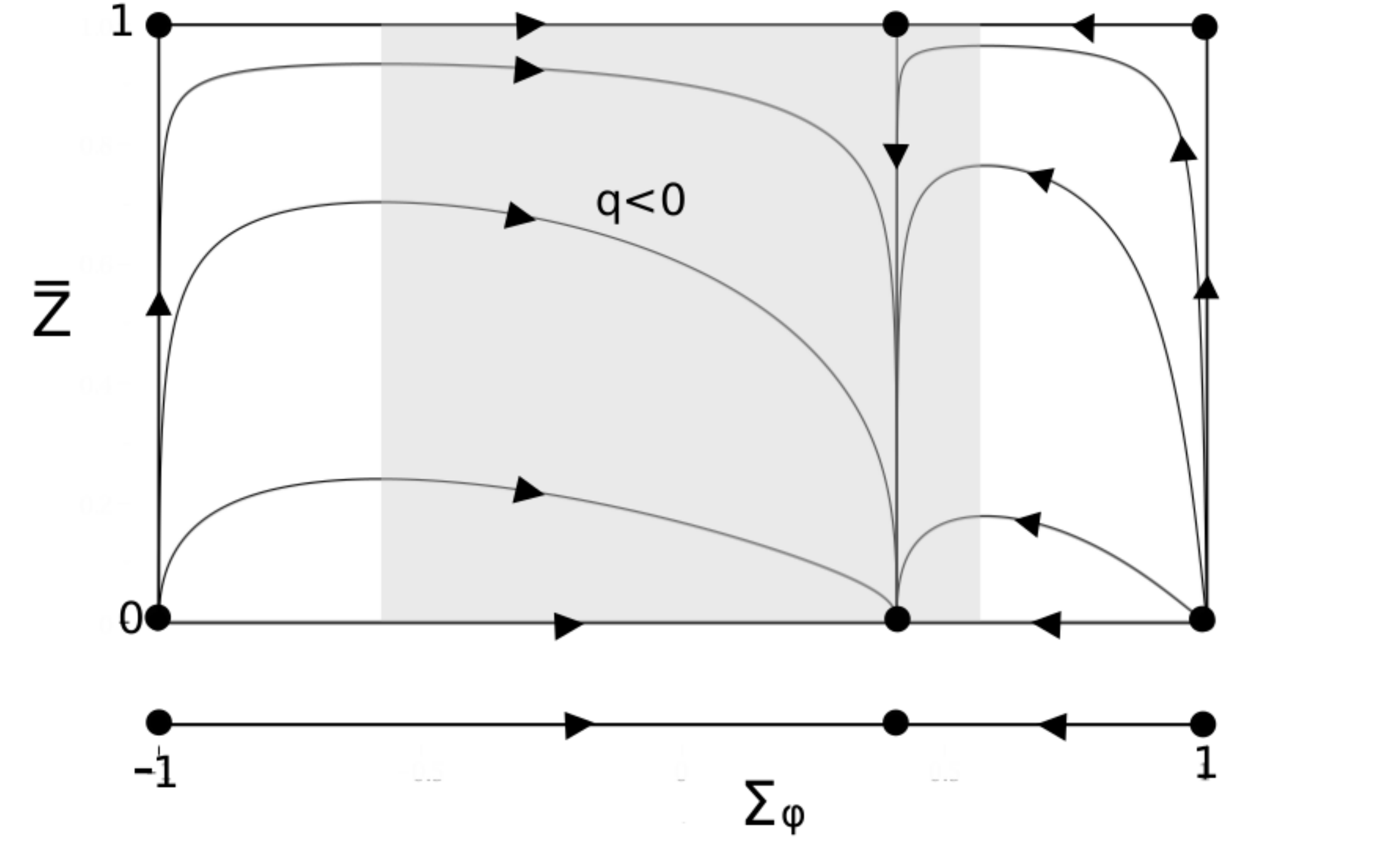}}  \hspace{0cm}
		\subfigure[$\lambda=1/\sqrt{2}$]{\label{}
			\includegraphics[width=0.45\textwidth]{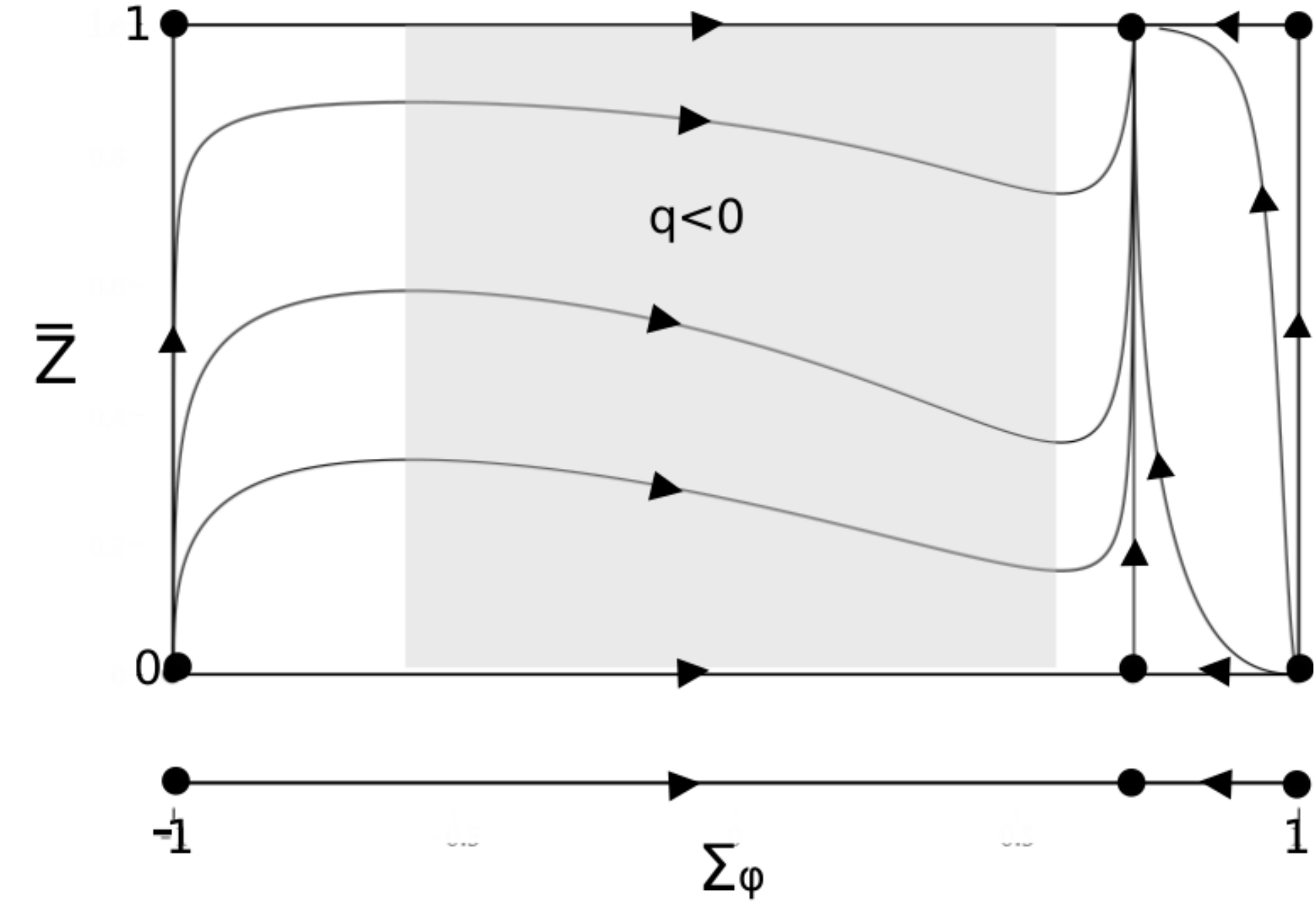}}
		\vspace{-.5cm}
	\end{center}
	\caption{Orbits and fixed points in the background state space ${\cal B}$ using $e$-fold time $N$,
	         with $0 \leq \lambda<1/\sqrt 3$, represented by $\lambda=1/\sqrt{6}$, in panel (a),
             and $1/\sqrt{3} < \lambda < 1$, represented by $\lambda=1/\sqrt{2}$, in panel (b).
             The shaded regions denote when $q<0$ and where $Z$ and $\bar{Z}$ are
             monotonically decreasing.
             }
	\label{FigEXP}
\end{figure}

The qualitative behaviour of typical and special orbits in the two
cases $0 \leq \lambda<1/\sqrt 3$ and $1/\sqrt{3} < \lambda < 1$
is illustrated in Fig.~\ref{FigEXP}.

Referring to Fig.~\ref{FigEXP}
we can summarize the asymptotic behaviour of the orbits in the background
state space $\cal B$ for $\lambda\neq1/\sqrt 3$ as follows. For evolution
into the future for all orbits with $\Sigma_{\varphi}^2<1$  we have
\begin{subequations} \label{B.future}
\begin{align}
\lim_{N\rightarrow \infty}\Sigma_{\varphi}&=\lambda,\\
\lim_{N\rightarrow \infty}{\bar Z}&=0,\quad \text {if}\quad 0 \leq \lambda < 1/\sqrt 3, \\
\lim_{N\rightarrow \infty}{\bar Z}&=1,\quad \text {if}\quad 1/\sqrt 3 < \lambda < 1.
\end{align}
\end{subequations}
Into the past all orbits with $\Sigma_{\varphi}\neq \lambda$  satisfy,
\begin{equation} \label{B.past}
\lim_{N\rightarrow -\infty}\Sigma_{\varphi}^2=1,\qquad
\lim_{N\rightarrow -\infty}{\bar Z}=0,
\end{equation}
while the special case $\Sigma_{\varphi}=\lambda$ results in
\begin{subequations} \label{B.asymp.special1}
\begin{align}
\lim_{N\rightarrow -\infty}{\bar Z} &= 1,
\quad \text {if}\quad 0 \leq \lambda< 1/\sqrt 3, \\
\lim_{N\rightarrow -\infty}{\bar Z} &= 0,
\quad \text {if}\quad 1/\sqrt 3 < \lambda < 1.
\end{align}
\end{subequations}
For the special case $\Sigma_{\varphi}=\pm1$ we have
\begin{equation} \label{B.asymp.special2}
\lim_{N\rightarrow \infty}{\bar Z}=1, \qquad
\lim_{N\rightarrow -\infty}{\bar Z}=0, \quad \text {for all}\,\, \lambda.
\end{equation}
In the next section we will use these results to determine the asymptotic
behaviour of the orbits in the full state space ${\cal B}\times{\cal P}$.

Before continuing we make some remarks relating our description of the
background dynamics to previous research.
The first paper to give a dynamical systems
analysis of a scalar field with exponential potential is Halliwell (1987)~\cite{hal87},
who used the background scalar field $\varphi$ and an exponential representation
of the cosmological scale factor as variables.\footnote{In his figure 2, the flat background
case is represented by the bold face hyperbola, and this corresponds to our background state
space on the $\Sigma_{\varphi}$-axis shown in our Figure 2 since the exponential scale
factor variable decouples in the flat case. The fixed point A in his figure 2 is the future
attractor, which corresponds to our fixed point with $\Sigma_{\varphi}=\lambda$.}
From our perspective, the limitation of his approach is that the state space is unbounded.
The bounded variable $\Sigma_\varphi$ that we are using is not new.
It has been introduced before in the context
of a scalar field with an exponential potential by Coley \emph{et al} (1997)~\cite{coletal97}
and by Copeland \emph{et al} (1998)~\cite{copetal98}. Both these references considered more
general background problems than the present one: Coley \emph{et al} studied anisotropic Bianchi
models and Copeland \emph{et al} added a perfect fluid with a linear equation of state to the
scalar field.  The variable $d\varphi/dN$, which is proportional to $\Sigma_\varphi$,
is now commonly used to describe scalar fields in cosmology.\footnote{See for example,
Urena-Lopez (2012)~\cite{ure12}, equation (2.3), Tsujikawa (2013)~\cite{tsu13},
equation (16) and Alho and Uggla (2015)~\cite{alhugg15b}, equation (8).}

Our state space ${\cal B}$ differs in an important way from~\cite{hal87, copetal98},
in that it also describes the dynamics of
the variable $Z=k^2/{\cal H}^2$, compactified to give ${\bar Z}$, whose
evolution is determined by $\Sigma_{\varphi}$ through the differential
equation~\eqref{barZN}, which leads to
the two dimensional state space in our figure 2. This yields a
complete description of the dynamics of
a scalar field with exponential potential in a flat FL background
needed to determine the evolution of perturbations
on this background, which is the main goal of this paper.
This representation of the dynamics, which to the best of our knowledge is new,
has several advantages. It shows directly in which
regions of the state space there is accelerated expansion (the shaded regions in figure~\ref{FigEXP},
in which $q=3\Sigma_{\varphi}^2-1<0$ and where thereby $Z$ and $\bar{Z}$
are monotonically decreasing), and highlights the physical interpretation
of the parameter $\lambda$ in the potential.
Referring to our figure 2, if $\lambda<1/\sqrt 3$
then all orbits eventually enter and remain in the region of
accelerated expansion and thus the
orbits describe universes
that undergo future accelerated expansion.  On the other hand if
$\lambda<1/\sqrt 3$ orbits that are past asymptotic to $\Sigma_{\varphi}=1$ do not
enter the shaded region and hence
never undergo accelerated expansion, while orbits that
are past asymptotic to $\Sigma_{\varphi}=-1$ enter and then leave the shaded
region and hence undergo transient accelerated expansion. We note that
the general orbits in Figure 2 have been given explicitly in
parametric form $H(u), \varphi(u),$ where $u$ is a parameter, by Salopek and
Bond (1990)~\cite{salbon90}, and analyzed qualitatively (see
equations (3.2) and their figure 1). Our state space picture shows this
behaviour directly.

\subsection{General  and special orbits in ${\cal B}\times{\cal P}$
\label{dynamics} }

Since the full state space has a product structure ${\cal S} = {\cal B}\times{\cal P}$,
the limits of $\Sigma_\varphi $ and $\bar{Z}$ as $N\rightarrow \pm\infty$
given by equations~\eqref{B.future}, \eqref{B.past}, \eqref{B.asymp.special1}
and~\eqref{B.asymp.special2}, which we derived in the background
state space ${\cal B}$ (see also Fig.~\ref{FigEXP}), are also valid in
${\cal B}\times{\cal P}$, and we can use these results to
determine the asymptotic behaviour of the orbits in the full state space.
We still limit the discussion to $0\leq \lambda <1$, and we also
exclude the bifurcation value $\lambda=1/\sqrt 3$.
The restrictions imposed in deriving~\eqref{B.future} and~\eqref{B.past} require that
we consider separately the \emph{generic orbits}
and the \emph{special orbits} that satisfy
$\Sigma_\varphi= \lambda$ and $\Sigma_\varphi=\pm 1$ with constant deceleration parameter
$q$. We refer to the case $\Sigma_\varphi= \lambda$ as the (interior) \emph{scale-invariant}
orbits,\footnote{The solutions corresponding to orbits with $\Sigma_\varphi=\lambda>0$ are invariant under
constant conformal scalings and admit a homothetic Killing vector field. The orbits with $\Sigma_\varphi=\lambda=0$
correspond to the one-parameter family of de Sitter solutions, where a constant scaling scales the
dimensional cosmological constant that is parametrizing the family of solutions.
Constant conformal scalings thereby result in the same solution when $\lambda>0$ while
they lead to a new member in the same family of solutions when $\lambda=0$.
The close connection between the two classes is illustrated by that both $\lambda >0$
and $\lambda =0$ yield constant values for dimensionless scalars such as $\Sigma_\varphi$
and $q$. For simplicity we will therefore also refer to $\lambda=0$ as belonging to the scale-invariant case.}
and the orbits with $\Sigma_\varphi=\pm 1$ as the \emph{massless scalar field} orbits
(although they also correspond to scale-invariant solutions).
On recalling that $q=3\Sigma_\varphi^2-1$, we have
\begin{equation} \label{classA,B}
q = 3\lambda^2-1\, \text{in the scale-invariant case}; \quad q=2 \, \text{in the massless scalar field case}.
\end{equation}
%

First as regards the \emph{generic orbits} it follows from
equations~\eqref{B.future} and~\eqref{B.past} that
\begin{itemize}
\item[Gi)] all generic orbits are past asymptotic to one of the fixed points $\mathrm{M}_{\pm}$,
which form the  past attractor of the dynamical system,
\item[Gii)] if  $\lambda < 1/\sqrt{3}$ then a two parameter family of
orbits is future asymptotic to the fixed
point $\mathrm{A}_{\lambda}$, which forms the future attractor; in addition
a one-parameter family of
orbits is future asymptotic to the fixed
point $\mathrm{S}_{\lambda}$,
\item[Giii)] if  $\lambda > 1/\sqrt{3}$ then all orbits are future
asymptotic to the heteroclinic cycle ${\cal H}_{\lambda}$, which forms the future attractor.
\end{itemize}
We give some details of the derivation of the result Gii) at the end of this section.

We now describe  the acceleration/deceleration properties of
these generic models.\footnote{The region of state space in which there is accelerated
expansion is a cylindrical shell in the state space ${\cal S}={\cal B}\times {\cal P}$,
given by the inequalities $-1/\sqrt 3<\Sigma_{\varphi}< 1/\sqrt 3.$ This region can be
visualized by rotating the shaded region in figure 2 about a line $\Sigma_{\varphi}=b<-1,$
as described at the beginning of section 4.1.}
Since $q=3\Sigma_\varphi^2-1$ we can draw
the following conclusions by referring to Fig.~\ref{FigEXP} and using
the results Gi)-Giii) above:
\begin{itemize}
\item[i)] If $\lambda<1/\sqrt 3$, then all models undergo \emph{future acceleration},

\item[ii)] If  $\lambda > 1/\sqrt{3}$ then all models undergo \emph{future deceleration}.
However, there are two subcases. Orbits that are past asymptotic to $\mathrm{M}_-$
yield models that undergo \emph{transient acceleration}, i.e.
they have a finite epoch during which $q<0$, see Fig.~\ref{FigEXP};
orbits that are past asymptotic to $\mathrm{M}_+$ yield models that
are forever decelerating, i.e. $q>0$ at all times.
\end{itemize}

\begin{figure}[ht!]
	\begin{center}
			\subfigure[$\lambda=1/\sqrt{6}$]{\label{}
			\includegraphics[width=0.45\textwidth]{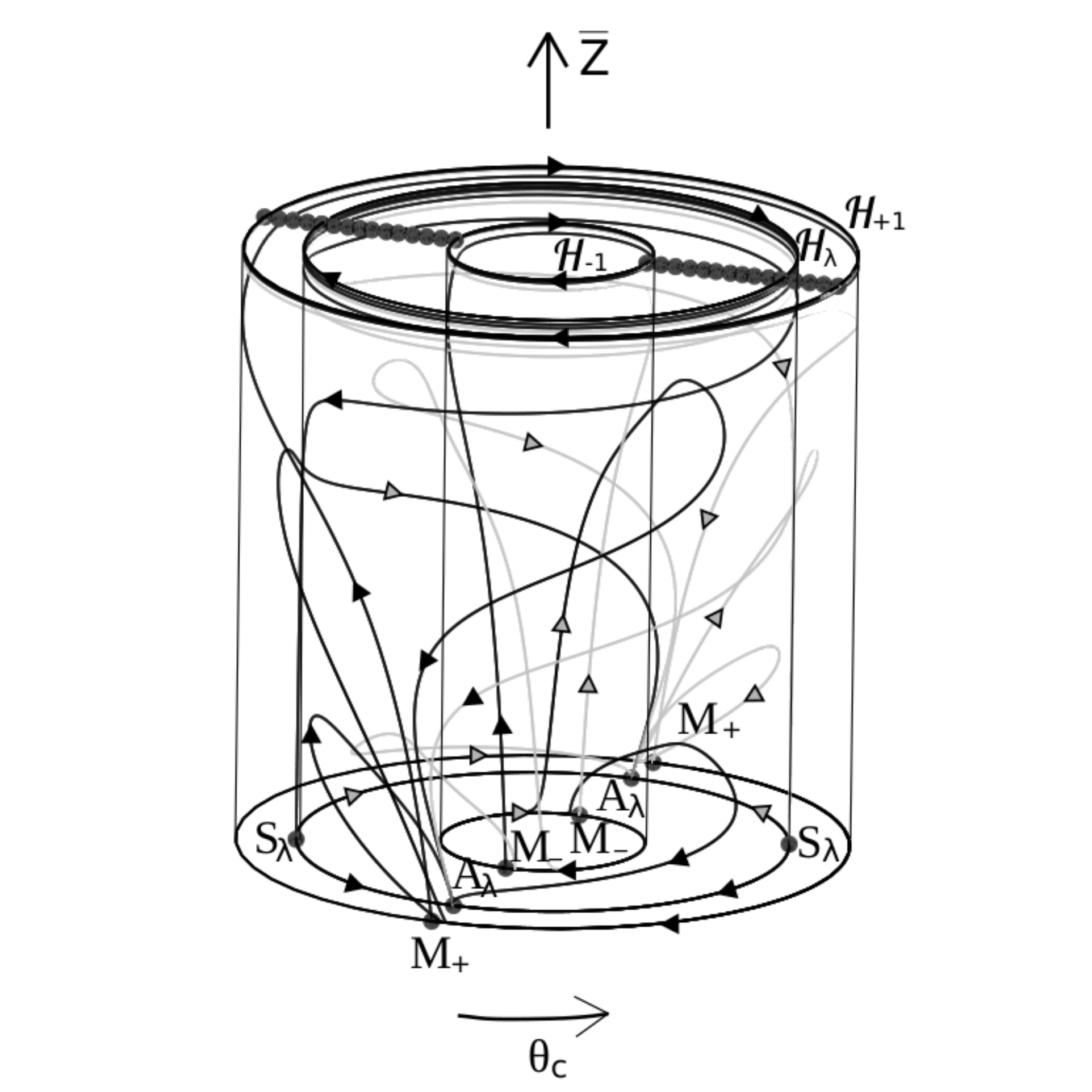}}
            \subfigure[$\lambda=1/\sqrt{2}$]{\label{}
			\includegraphics[width=0.45\textwidth]{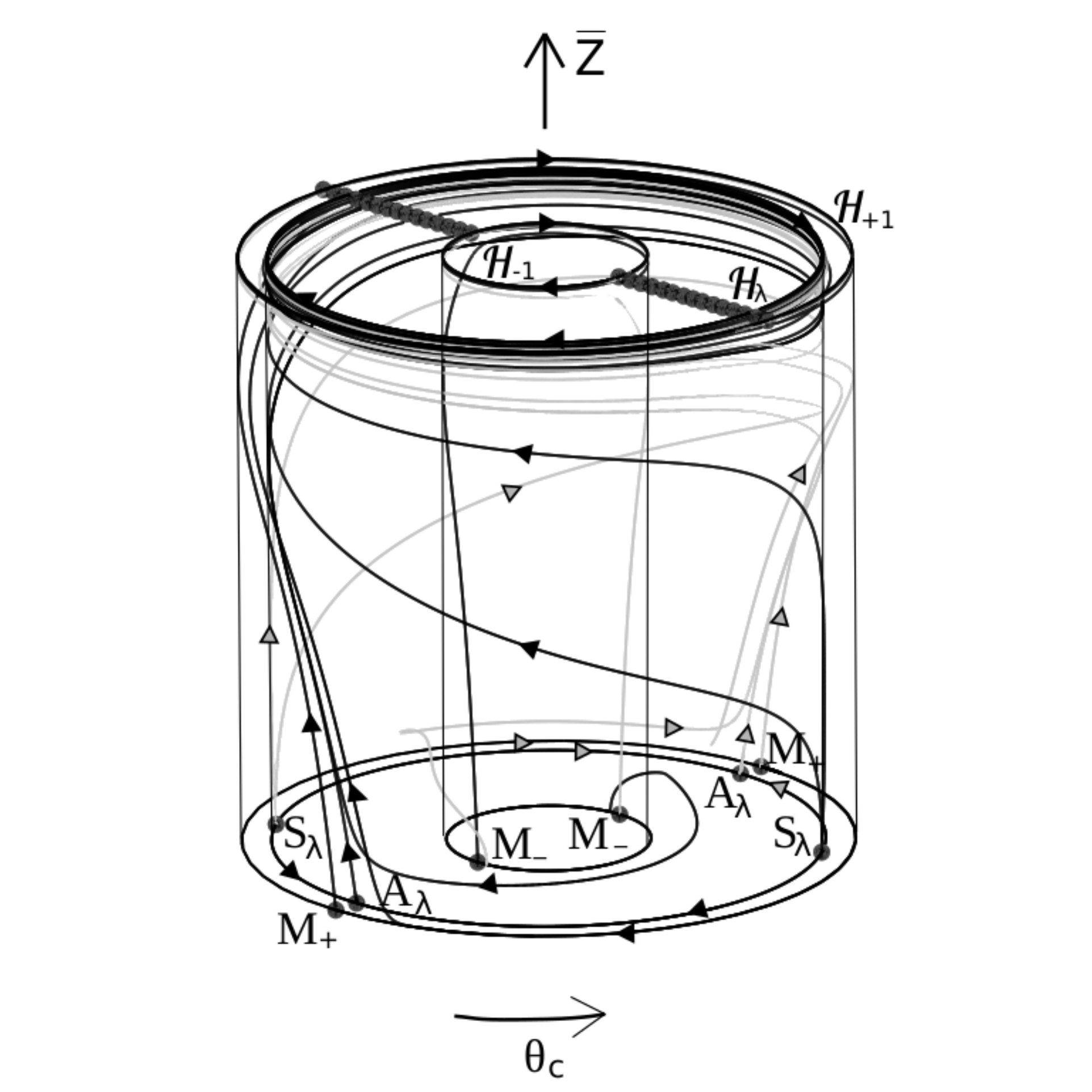}}
		\vspace{-0.5cm}
	\end{center}
	\caption{Representative orbits on the state space
             ${\cal S} = {\cal B}\times {\cal P}$ as described in results Gi)-Giii),
             with $0 \leq \lambda<1/\sqrt 3$, illustrated by $\lambda=1/\sqrt{6}$, in panel (a), and
             $1/\sqrt{3} < \lambda < 1$, illustrated by $\lambda=1/\sqrt{2}$, in panel (b).}
           	\label{FigEXPPERT_INT}
\end{figure}

Fig.~\ref{FigEXPPERT_INT} shows examples of generic orbits in the two cases
$0 \leq \lambda < 1/\sqrt{3}$ and $1/\sqrt{3} < \lambda < 1$.

Before continuing we make a remark concerning the physical interpretation
of Figure~\ref{FigEXPPERT_INT}. It is helpful to think how the
orbits are generated numerically. One chooses initial values for
$\Sigma_{\varphi}, {\bar Z}$ and $\theta$ and then integrates numerically
forwards and backwards to generate an orbit.\footnote{The usual way
to numerically integrate the perturbation equations
for scalar fields is to first integrate the background equations, and then use the output
to integrate the perturbed Klein-Gordon equation to determine the
real and imaginary parts of the
Fourier transform of the scalar field perturbation, for a given wave number $k$.
One is thereby treating the perturbed Klein-Gordon equation as a non-autonomous
differential equation, in contrast to our dynamical systems approach.
See for example Huston and Malik (2009)~\cite{husmal09} (this paper
also covers second order perturbations) and Martin and
Ringeval (2006)~\cite{marrin06} (see section 4.1).}
One can use the initial value of ${\bar Z}$
in conjunction with equation~\eqref{bar.Z} to determine the wave number $k$
in terms of the value of the Hubble scalar
$H_0$ at the initial time $a_0=1, N_0=0$:
\begin{equation}
{\bar Z}_0=\frac{k^2}{k^2 + {\cal H}_0^2}=\frac{k^2}{k^2 + H_0^2}.
\end{equation}
In this way one can generate a family of orbits corresponding to different
wave numbers.

\begin{figure}[ht!]
	\begin{center}
		\subfigure[$\lambda=1/\sqrt{6}$]{\label{}
		\includegraphics[width=0.45\textwidth]{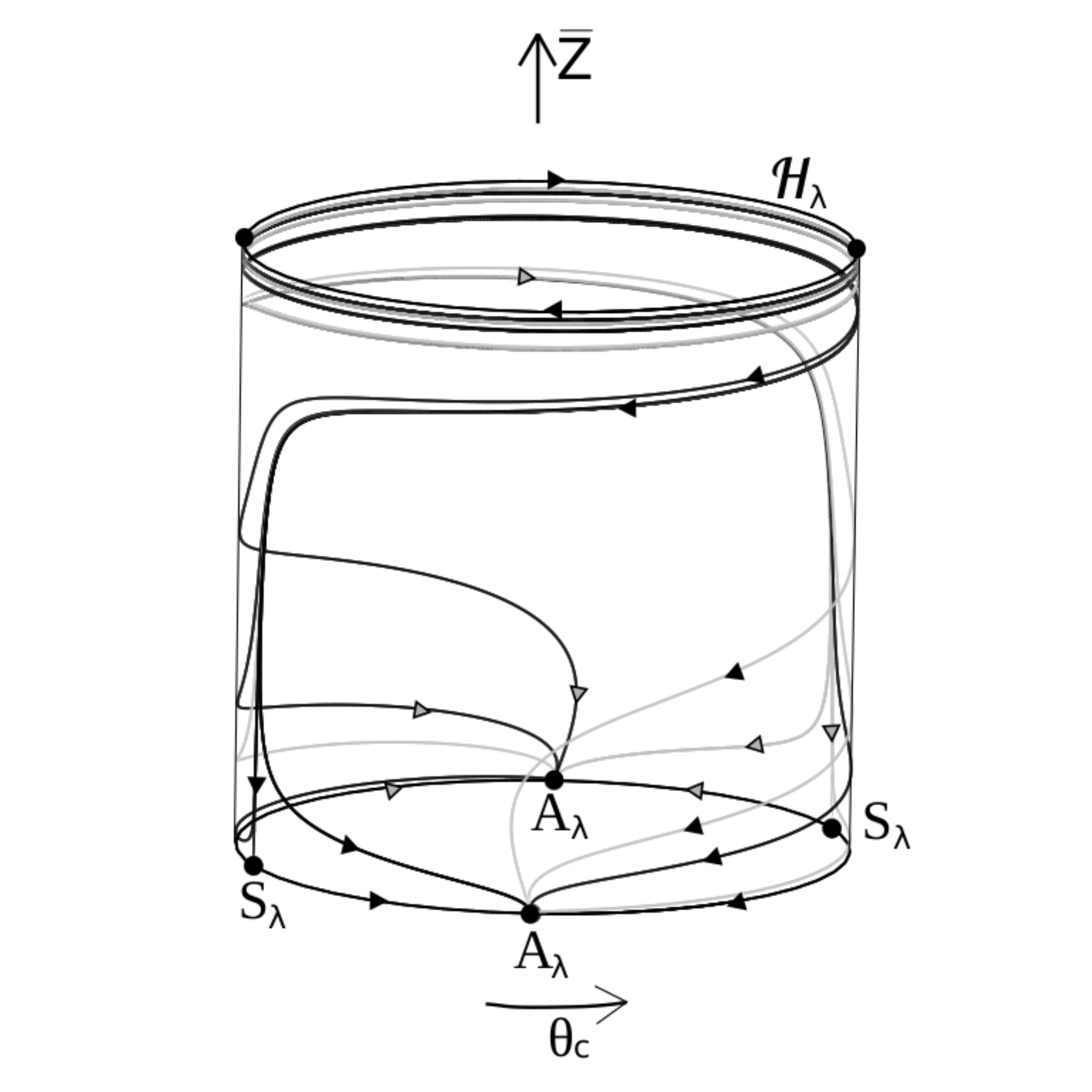}} 
		\subfigure[$\lambda=1/\sqrt{2}$]{\label{}
			\includegraphics[width=0.48\textwidth]{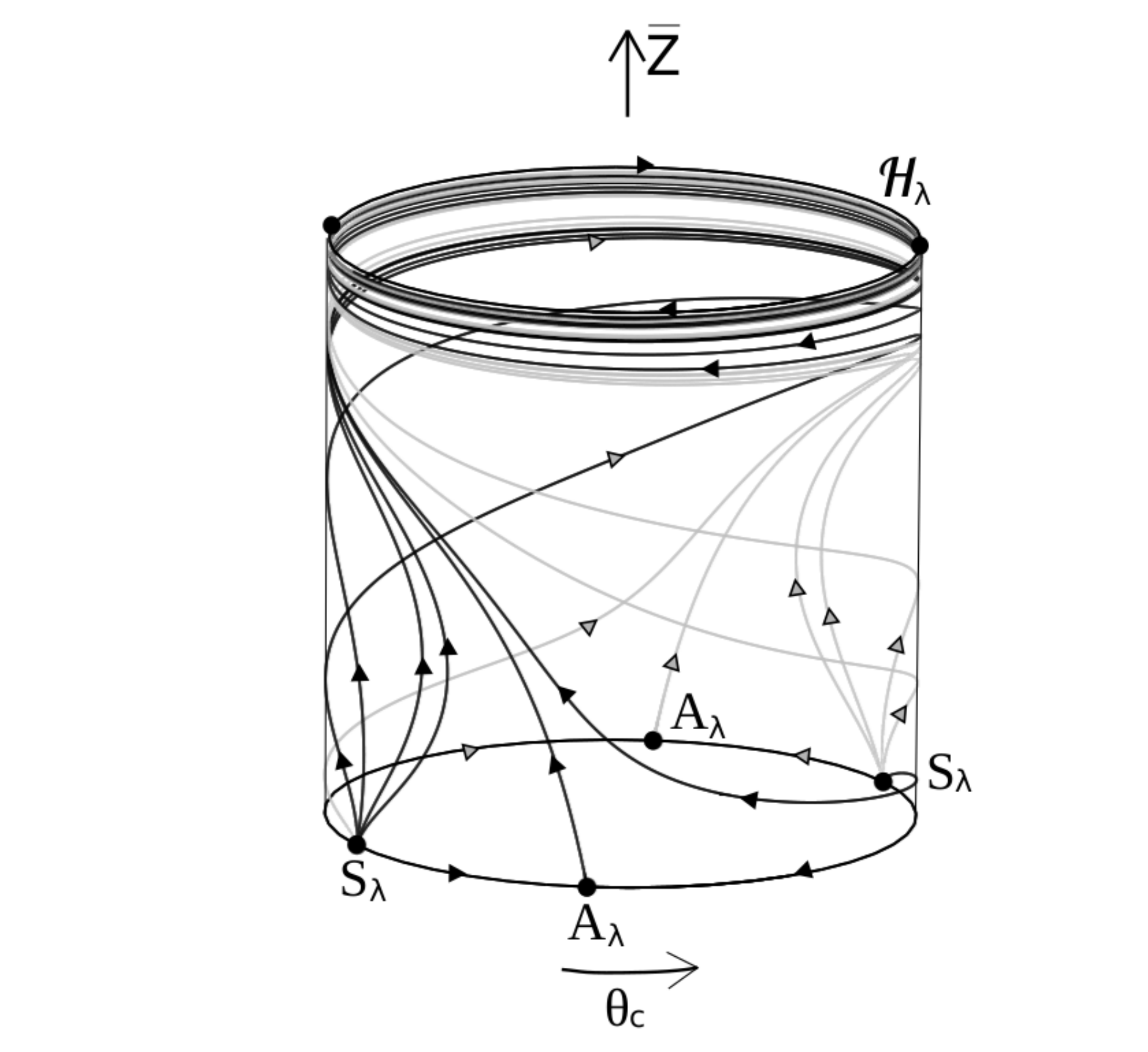}} \\
		\vspace{-0.5cm}
	\end{center}
	\caption{Representative orbits on the invariant set $\Sigma_\varphi = \lambda$:
	Ai) for which $0 \leq \lambda<1/\sqrt 3$ is illustrated by $\lambda=1/\sqrt{6}$
    in panel (a), while Aii) for which $1/\sqrt{3} < \lambda < 1$ is illustrated by
    $\lambda=1/\sqrt{2}$ in panel (b).}
	\label{FigEXPyZ}
\end{figure}

Next, as regards the interior \emph{scale-invariant orbits} ($\Sigma_\varphi= \lambda < 1$) it follows
from equation~\eqref{B.asymp.special1} that
\begin{itemize}
\item[Ai)] if  $0 \leq \lambda < 1/\sqrt{3}$ (i.e. $q<0$)
then a scale-invariant orbit is past asymptotic to the
heteroclinic cycle ${\cal H}_{\lambda}$ and is future asymptotic
to the fixed point $\mathrm{A}_{\lambda}$,
apart from one exceptional orbit that is future asymptotic
to the fixed point $\mathrm{S}_{\lambda}$,
\item[Aii)] if $1/\sqrt{3} < \lambda < 1$ (i.e. $q>0$)
then a scale-invariant orbit is future asymptotic to the
heteroclinic cycle ${\cal H}_{\lambda}$ and is past asymptotic to the
fixed point $\mathrm{S}_{\lambda}$, apart from one exceptional orbit that is past
asymptotic to the fixed point $\mathrm{A}_{\lambda}$.
 \end{itemize}
Representative scale-invariant orbits are depicted in Fig.~\ref{FigEXPyZ} (recall that there are two copies
of fixed points and orbits when using $\theta$ instead of $y$, related to each other by
$\theta \rightarrow \theta + \pi$).

\begin{figure}[ht!]
	\begin{center}
			\includegraphics[width=0.45\textwidth]{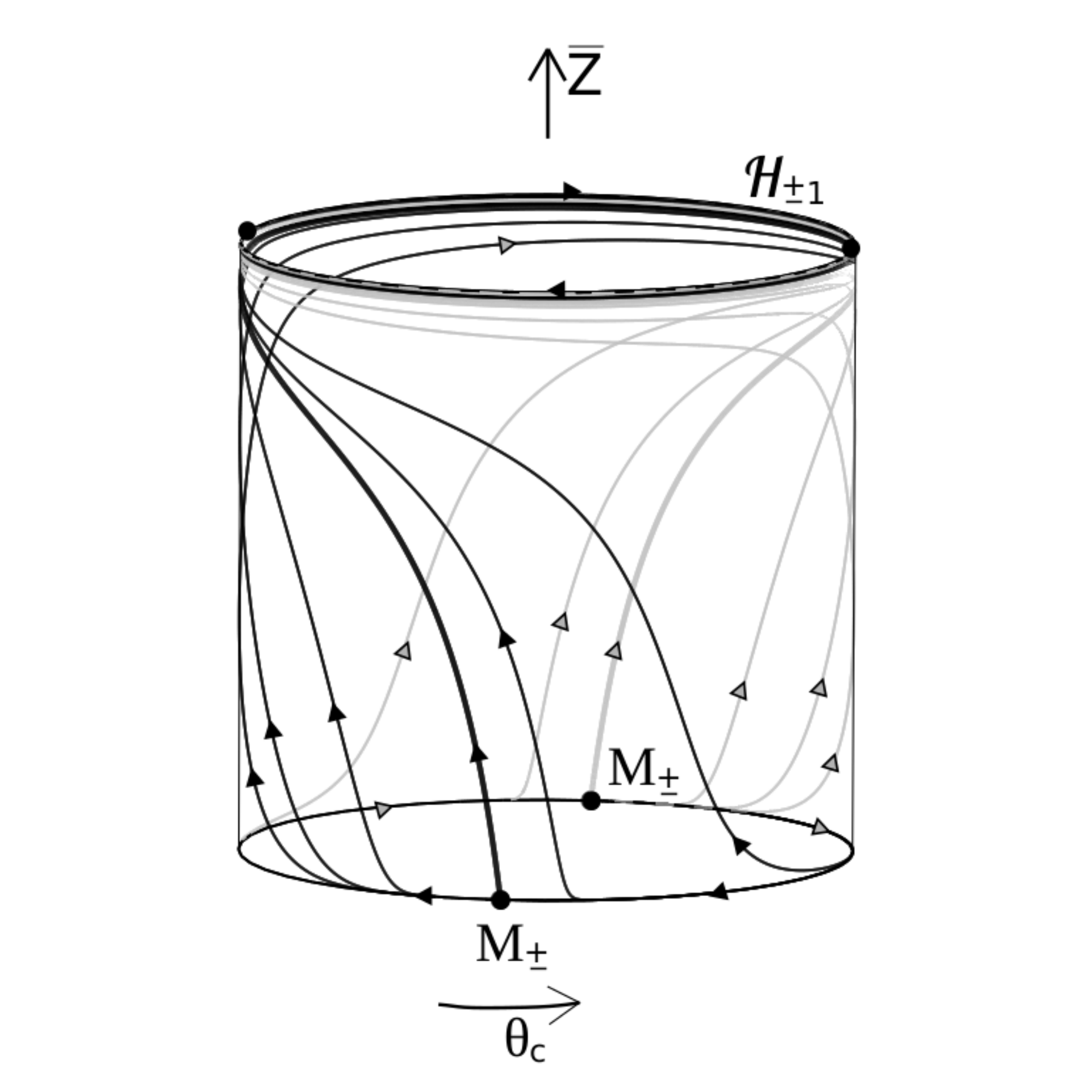}
		\vspace{-0.5cm}
	\end{center}
	\caption{Representative orbits on the massless scalar field boundary for
	any value of $\lambda$
    (the orbits on the two components $\Sigma_\varphi = - 1$
    and $\Sigma_\varphi = +1$ are identical).}
	\label{FigEXPyZMassless}
\end{figure}

Finally, as regards the \emph{massless scalar field orbits} ($\Sigma_\varphi= \pm 1$) it follows
from equations~\eqref{B.asymp.special2} that such
an orbit is past asymptotic to one of the fixed points $\mathrm{M}_{\pm}$,
and is future asymptotic to one of the heteroclinic cycles ${\cal H}_{\pm 1}$.
Representative massless scalar field orbits are depicted in Fig.~\ref{FigEXPyZMassless}.

In deriving the result Gii), the limits~\eqref{B.future} in ${\cal B}$
imply that $\lim_{N\rightarrow \infty}\Sigma_{\varphi}=\lambda,$
and $\lim_{N\rightarrow \infty}{\bar Z}=0$ in the state space
${\cal S} = {\cal B}\times{\cal P}$. It follows that the orbits in question are future
asymptotic to the one dimensional invariant set  $\Sigma_{\varphi}=\lambda,
\bar{Z}=0$ (a circle). Since this circle contains the two hyperbolic fixed points $\mathrm{A}_{\lambda}$
and $\mathrm{S}_{\lambda}$, the orbits in question must be future asymptotic
to one of these fixed points. The final step is to note that in this case
$\mathrm{A}_{\lambda}$ is a sink and $\mathrm{S}_{\lambda}$ is a saddle with a two dimensional
stable manifold.

The local stability properties of $\mathrm{A}_{\lambda}$ and $\mathrm{S}_{\lambda}$
also yield the local stability properties for Ai) and Aii):
in Ai) $\mathrm{A}_{\lambda}$ is a sink and
$\mathrm{S}_{\lambda}$ is a saddle on the interior scale-invariant set $\Sigma_\varphi=\lambda$,
while in Aii) $\mathrm{S}_{\lambda}$ is a source and $\mathrm{A}_{\lambda}$ is a
saddle, see Fig.~\ref{FigEXPyZ}.

\subsection{Asymptotic expansions}\label{subsec:asympt}

In this subsection we will give asymptotic expansions for $\Sigma_\varphi$,
$Z$, $y$ and $\varphi_\mathrm{c}$ in terms of the $e$-fold time $N$ in the
vicinity of the fixed points in the long wavelength boundary set $Z = 0$.
When it comes to $\varphi_\mathrm{c}$, we will focus on the future accelerating
case $0 \leq \lambda < 1/\sqrt{3}$, for which we need $\varphi_\mathrm{c}$ in
section~\ref{sec:cons}.

\subsubsection*{Asymptotics  for the fixed point $\mathrm{A}_\lambda$}

We have shown that for generic orbits with $0\leq\lambda<1/\sqrt{3}$
the future attractor is the fixed point $\mathrm{A}_\lambda$:
$\Sigma_{\varphi}=\lambda$, $Z=0$, $y = 0$, with $q = 3\lambda^2 - 1 <0$.
Solving the evolution equations~\eqref{dynsystotN}
to leading order in the vicinity of this fixed point
yields the following expressions:
\begin{subequations}\label{Al}
\begin{align}
\Sigma_{\varphi} &\approx\lambda + C_1 e^{-3(1-\lambda^2)N},\\
Z &\approx C_2 e^{-2(1-3\lambda^2)N},\\
y &\approx C_3e^{-3(1-\lambda^2)N} -
\frac{C_2}{1+3\lambda^2}e^{-2(1-3\lambda^2)N}, \label{Al.3}
\end{align}
\end{subequations}
as $N\rightarrow +\infty$, where $C_1$, $C_2\geq 0$, and $C_3$ are constants,
where $C_2 > 0$ when $k^2 >0$. The scale-invariant orbits
are given by $C_1=0$. It follows from~\eqref{Al.3}
and~\eqref{y.det.f} that for the orbits,
which are all converging to $\mathrm{A}_\lambda$,
\begin{equation}
\varphi_\mathrm{c} \approx C_\varphi^+
\left(1 - \frac{C_3}{3(1-\lambda^2)}e^{-3(1-\lambda^2)N} + \frac{C_2}{2(1-9\lambda^4)}e^{-2(1-3\lambda^2)N}\right),
\end{equation}
where $\lim_{N\rightarrow +\infty}\varphi_\mathrm{c} = C_\varphi^+$
when $0\leq\lambda<1/\sqrt{3}$.
For future decelerating models ($1/\sqrt{3} < \lambda < 1$)
there is a unique orbit in the invariant set $\Sigma_\varphi = \lambda$,
characterized by $C_1=C_3=0$ that is past asymptotic to $A_{\lambda}$.

\subsubsection*{Asymptotics  for the fixed point $\mathrm{S}_\lambda$}

The linear solution describing the solutions in the vicinity of
$\mathrm{S}_\lambda$ is given by
\begin{subequations}\label{linS}
\begin{align}
\Sigma_{\varphi} &= \lambda + C_1e^{-3(1-\lambda^2)N},\\
Z &= C_2 e^{-2(1-3\lambda^2)N},\\
y &= -3(1 - \lambda^2) + 3{\lambda}C_1 e^{-3(1-\lambda^2)N} + \frac{C_2}{5 - 9\lambda^2}e^{-2(1-3\lambda^2)N}
+ C_3e^{3(1 - \lambda^2)N},\label{Sy}
\end{align}
\end{subequations}
where again $C_2 > 0$ when $k^2 >0$. This implies that for the future accelerating models
$0\leq\lambda<1/\sqrt{3}$ there is a one-parameter
family of orbits that are future asymptotic to the hyperbolic saddle point
$\mathrm{S}_\lambda$, characterized by $C_3=0$, where the orbit with $C_1=0$
lies in the interior scale-invariant set $\Sigma_{\varphi}=\lambda$.
In this case
\begin{equation}\label{varphiS}
\varphi_\mathrm{c} \approx C_\varphi^+ e^{-3(1-\lambda^2)N}
\left(1 - \frac{\lambda C_1}{1-\lambda^2}e^{-3(1-\lambda^2)N} -
\frac{C_2}{2(1-3\lambda^2)(5-9\lambda^2)}e^{-2(1-3\lambda^2)N}\right),
\end{equation}
where $\lim_{N\rightarrow +\infty}\varphi_\mathrm{c} = 0$.
For the future decelerating models with $1/\sqrt{3} < \lambda < 1$ there is a one-parameter
family of orbits that are past asymptotic to
$\mathrm{S}_\lambda$ in the interior scale-invariant set
$\Sigma_\varphi = \lambda$, characterized by $C_1=0$.

\subsubsection*{Asymptotics  for the fixed points $\mathrm{M}_\pm$}

We have shown that for generic orbits the past attractor
is the pair of fixed points $\mathrm{M}_\pm$:
$\Sigma_{\varphi}=\pm 1$, $\bar{Z} = Z = 0$, $y = 0$.
Solving the evolution equations~\eqref{dynsystotN}
to leading order in the vicinity of these fixed points results in
\begin{subequations}\label{AsymM}
\begin{align}
\Sigma_{\varphi} &\approx \pm (1 - C_1 e^{6(1\mp\lambda)N}),\\
Z &\approx C_2e^{4N},\\
y &\approx  \frac{C_3}{1+C_3 N} - \frac{C_2}{4}e^{4N} - 6(1\mp\lambda)C_1 e^{6(1\mp\lambda)N}, \label{AsymM.3}
\end{align}
\end{subequations}
as $N\rightarrow-\infty$, where $C_1\geq 0$, $C_2 \geq 0$, $C_3$ are constants,
with $C_2 \propto k^2$ where $C_2 > 0$ when $k^2 >0$, and where we have used the
translation freedom in $N$ to set a constant $A$ in a term $C_3/(A + C_3 N)$ to one.
Note that $C_3=0$ describes the two-dimensional unstable manifolds of
the fixed points $\mathrm{M}_\pm$, while the power law approach to
zero in the $C_3$ term is due to presence of a zero eigenvalue and
resulting centre manifold. It follows from~\eqref{AsymM.3} and~\eqref{y.det.f} that
to leading order
\begin{equation}\label{varphiMpm}
\varphi_\mathrm{c} \approx C_\varphi^-(1 + C_3N)
\left(1 - \frac{C_2}{16}e^{4N} - C_1 e^{6(1\mp\lambda)N}\right),
\end{equation}
where $\lim_{N\rightarrow - \infty}\varphi_\mathrm{c} = C_\varphi^-$
when $C_3 = 0$, while $\varphi_\mathrm{c}$ becomes unbounded toward the past
when $C_3 \neq 0$.


%
%
%
%

\section{Explicit solutions with constant deceleration parameter \label{bessel}}

The Klein-Gordon equation~\eqref{FourierKG2scalarsexp}
for a perturbed scalar field with exponential potential
has been solved explicitly for $\varphi_{\mathrm c}$ subject to
the assumption that the scale factor $a$ has a power law dependence
on conformal time, and thereby a constant deceleration parameter $q$
(see footnote~\ref{fn:2} for references).
In this case it is convenient to introduce a new variable
\begin{equation} \label{def.v}
v=a\varphi_{\mathrm c},
\end{equation}
and to use conformal time $\eta$ instead of $e$-fold time $N$.\footnote{In
making the transition from $N$ to $\eta$ we use the relations
$\partial_{\eta}={\cal H}\partial_N$,
$\partial_{\eta}^2={\cal H}^2(\partial_N^2-q\partial_N)$,
and equation~\eqref{prop.eta}.}

Recall that $\varphi_{\mathrm c}$ is the complex Fourier
coefficient of the scalar field perturbation in the uniform curvature gauge.
Note that we have dropped the index $k$ on
$\varphi_{{\mathrm c},k}$, and where we now
similarly drop the index $k$ on $v_k$. Making the above changes transform
equation~\eqref{FourierKG2scalarsexp} into the following Bessel equation
for the function $v$:\footnote{Since $q=3\Sigma_\varphi^2-1$ by~\eqref{q.phi},
$q'=0\implies \Sigma_\varphi'=0$, which gives
$(1-\Sigma_\varphi^2)(\lambda-\Sigma_\varphi)=0$
using~\eqref{Sigmaeq}. These results reduce equation~\eqref{FourierKG2scalarsexp}
to $\partial_N^2 \varphi_{\mathrm c} +(2-q)\partial_N\varphi_{\mathrm c}
+k^2{\cal H}^{-2}\varphi_{\mathrm c}=0$.  The next stage in the derivation is
$\partial_\eta^2 \varphi_{\mathrm c} +2{\cal H}\partial_\eta\varphi_{\mathrm c}
+k^2\varphi_{\mathrm c}=0$.}
\begin{equation} \label{MS3}
\partial_\eta^2\,v + \left(k^2-(\nu^2-\sfrac14){\eta}^{-2}\right)v = 0,
\end{equation}
where the index $\nu$ is determined by $q$ according to
\begin{equation} \label{index.q}
\nu = \frac{2-q}{2|q|}.
\end{equation}

In analyzing the solutions we need the following properties of
conformal time that follow from the assumption that $q$ is constant:
\begin{equation} \label{prop.eta}
{\cal H}\eta =q^{-1}, \qquad \eta/\eta_0=e^{qN}, \qquad  {\cal H} = {\cal H}_0 e^{-qN}.
\end{equation}
Since we are considering
expanding models (${\cal H}>0$), $\eta$ has the same
sign as $q$. If $q<0$, then $-\infty<\eta<0$ and
the limit $\eta\rightarrow 0^{-}$ describes late times
since $N\rightarrow \infty$ as $\eta\rightarrow 0^{-}$, while the limit
$\eta\rightarrow -\infty$ describes early times
since $N\rightarrow -\infty$ as $\eta\rightarrow -\infty$.
If $q>0$, then $0<\eta<\infty$ where
the limit $\eta\rightarrow 0^{+}$ describes early times
since $N\rightarrow -\infty$ as $\eta\rightarrow 0^{+}$,
while the limit $\eta\rightarrow \infty$ describes late times
since $N\rightarrow \infty$ as $\eta\rightarrow \infty$.

The general solution of Bessel's equation~\eqref{MS3} is
given in terms of Bessel functions by
\begin{equation} \label{DS2}
v = \sqrt{k|\eta|}\left(C_+J_{\nu}(k|\eta|) + C_-Y_{\nu}(k|\eta|)   \right),
\end{equation}
where $C_+$ and $C_-$ are complex constants that depend on $k$,
and $v=a\varphi_{\mathrm c}$.

The reduced dynamical variable $y$ is defined in equation~\eqref{theta.y}
by $y=f'/f$, where $f$ is a real function, either the real or imaginary part of
$\varphi_{\mathrm c}$.
The function $y=f'/f$ can be expressed in terms of
the corresponding function $v=af$ according to:
\begin{equation} \label{DS1}
y = \frac{f'}{f} = \frac{v'}{v} - 1 = \frac{\partial_\eta v}{{\cal H} v} - 1 =
\frac{q\,\eta\,\partial_\eta v}{v} -1,
\end{equation}
where we have used the first equation in~\eqref{prop.eta}.
This expression for $y$, with $v$ being any real function contained in
the general solution~\eqref{DS2},
describes the one-parameter\footnote{This
expression for $y$ can be written in terms of
only one parameter, either $C_+/C_-$ or $C_-/C_+$, without loss of generality,
when $C_-\neq 0$ or $C_+\neq 0$, respectively.}
family of orbits of the dynamical system for which $q$ is constant, in
other words the scale-invariant and massless scalar field orbits.
In order to complete the identification we
need to determine the asymptotic form of $y$, given by~\eqref{DS2} and~\eqref{DS1}
as $\eta\rightarrow 0$ and $\eta\rightarrow\pm \infty$.

The asymptotic form of the Bessel functions is
\begin{subequations} \label{bessel.asymp}
\begin{equation}  \label{bessel.asymp.1}
J_{\nu}(z)\sim z^{\nu}, \quad \text{for} \quad \nu\geq0
\quad\text{as}\quad z\rightarrow 0,
\end{equation}
\begin{equation}  \label{bessel.asymp.2}
Y_{\nu}(z)\sim z^{-\nu}, \quad \text{for} \quad \nu>0, \quad
Y_{0}(z)\sim \ln z, \quad\text{as}\quad z\rightarrow 0,
\end{equation}
\end{subequations}
and
\begin{equation} \label{Bessel.infty}
J_{\nu}(z) +iY_{\nu}(z)\approx
\sqrt{\frac{2}{\pi z}}\exp\left[ i\left(z-\sfrac14(2\nu+1)\pi\right)\right],
\quad\text{as}\quad z\rightarrow \infty,
\end{equation}
where $\sim$ means proportional to, i.e., we are dropping constant factors
that depend on $\nu$.

It follows from~\eqref{DS2} and~\eqref{bessel.asymp} that for the scale-invariant orbits
($\Sigma_\varphi = \lambda < 1$) there are two
cases as $|\eta|\rightarrow 0$. First the general case $C_-\neq 0$
in which $Y_{\nu}$ will be the dominant term, and a special case
$C_-=0$  in which $Y_{\nu}$  drops out making
$J_{\nu}$ the dominant term. We now use~\eqref{DS2}, \eqref{DS1},
and~\eqref{bessel.asymp}
to calculate the asymptotic form of $y$:
\begin{subequations}
\begin{align}
C_-\neq 0 \implies\, v \sim |\eta|^{1/2-\nu}
\Longrightarrow\, y\rightarrow q(\sfrac12-\nu) -1,
\quad \text{as}\quad |\eta|\rightarrow 0,\\
C_-=0 \implies\, v \sim |\eta|^{1/2+\nu}
\Longrightarrow\, y\rightarrow q(\sfrac12+\nu) -1,
\quad \text{as}\quad |\eta|\rightarrow 0.
\end{align}
\end{subequations}
It follows from~\eqref{MS3} that $q\nu=-\sfrac12(2-q)$,\, if $q<0$,
and $q\nu=\sfrac12(2-q)$,\, if $q>0$. The final results when
$|\eta|\rightarrow 0$ are:
\begin{subequations}
\begin{alignat}{9}
\text{If} \,\,& q<0, &\,\, \text{then}
\,\, C_- &\neq 0\,\, &\Longrightarrow\, y &\approx 0,
\quad &C_- &= 0\,\, &\Longrightarrow\, y &\approx -(2-q), \\
\text{If} \,\, & q>0, &\,\, \text{then}
\,\, C_- &\neq 0\,\, &\Longrightarrow\, y &\approx -(2-q),
\quad &C_- &= 0\,\, &\Longrightarrow\, y &\approx 0.
\end{alignat}
\end{subequations}
These limiting values are consistent with the fixed points in
the interior scale-invariant set $\Sigma_{\varphi}=\lambda$ that are given by
$\mathrm{A}_\lambda$: $y = 0$,
and $\mathrm{S}_\lambda$: $y = -3(1-\lambda^2)  = -(2-q)$.
If $q<0$ this gives the evolution into the future and for $q>0$
the evolution into the past.

In the limit $|\eta|\rightarrow \infty$, it follows
from equations~\eqref{DS2} and~\eqref{Bessel.infty}
that $v\sim \exp(\pm ik|\eta|)$ is sinusoidal.
Thus if $q<0$ ($q>0$) the evolution
into the past (future) is sinusoidal in terms of conformal time.\footnote{It
is of interest that in the case $q<0$ the initial condition for $v$ that
leads to power law inflation and the derivation of the power spectrum for the perturbations
is $v=(1\sqrt{2k}) \exp(-ik\eta)$, as $\eta\rightarrow -\infty$.
See, for example Durrer (2008)~\cite{dur08},  pages 112-114, in particular
equation (3.47).}
In our dynamical systems framework, however, this asymptotic behaviour
is described by the shadowing of the heteroclinic cycle ${\cal H}_{\lambda}$,
which is infinitely repetitive but not sinusoidal.

We have thus confirmed the asymptotic properties of the scale-invariant orbits:
\begin{itemize}
\item[Ai)] Accelerating models ($q<0$).\newline
Orbits with $C_-\neq 0$ evolve from ${\cal H}_{\lambda}$ to $\mathrm{A}_\lambda$.\newline
The exceptional orbit with $C_-= 0$ evolves from ${\cal H}_{\lambda}$ to
$\mathrm{S}_\lambda$.
\item[Aii)] Decelerating models ($q>0$).\newline
Orbits with $C_-\neq 0$ evolve from $\mathrm{S}_\lambda$ to
${\cal H}_{\lambda}$.\newline
The exceptional orbit with $C_-= 0$ evolves from the fixed point
$\mathrm{A}_\lambda$ to ${\cal H}_{\lambda}$.
\end{itemize}
%


For the massless scalar field orbits ($q=2$) we have $\eta>0$ and the expression
for $y$ is given by~\eqref{DS2} and~\eqref{DS1} with $\nu=0$.
It follows from~\eqref{bessel.asymp.2} that
\begin{subequations}
\begin{alignat}{8}
C_- &\neq 0 &\implies\, & v \sim\sqrt{\eta}\ln\eta\,\,
&\Longrightarrow\, &y &\rightarrow 0,
\quad &\text{as}\quad &\eta\rightarrow 0,\\
C_- &= 0 &\implies\, & v \sim \sqrt{\eta}\,\,
&\Longrightarrow\, &y &\rightarrow 0,
\quad &\text{as}\quad &\eta\rightarrow 0,
\end{alignat}
\end{subequations}
the difference being that $y$ tends to zero logarithmically
when $C_- \neq 0,\,(y\approx 2/\ln{\eta})$.\footnote{The
logarithmic approach to zero in $\eta$ is due to presence of a zero eigenvalue and
resulting centre manifold for the fixed points $\mathrm{M}_{\pm}$.}
Thus the massless scalar field orbits are past asymptotic
to one of the fixed points $\mathrm{M}_{\pm}$.
As $\eta\rightarrow \infty$, equation~\eqref{Bessel.infty} implies that
$y$ is oscillatory. Since the massless scalar field orbits
satisfy $\Sigma_{\varphi} =\pm1$ this confirms that they are future asymptotic to
one of the heteroclinic cycles ${\cal H}_{\pm}$.

\section{The comoving and uniform density curvature perturbations}\label{sec:cons}

It is generally believed that the comoving curvature perturbation,
${\cal R}$, and the uniform density curvature perturbation, $\zeta$, are
conserved quantities for long wavelength adiabatic perturbations,\footnote{For
the definition of these quantities in terms of gauge invariants,
and a brief history with other references, see for example Uggla and
Wainwright (2019)~\cite{uggwai19b}, section 4.}
a feature that plays an important role in inflationary cosmology.
This heuristic statement means that
in a time interval during which $k/{\cal H}\ll 1$  these gauge
invariants are approximately constant and in addition are
approximately equal in value. This result applies to a barotropic perfect
fluid and to a minimally coupled scalar field with arbitrary potential.
However, it has been pointed out by
Romano \emph{et al} (2016)~\cite{rometal16}
that it breaks down toward the future in the special case of a scalar field
with a constant potential, which is said to describe
so-called ultra-slow roll inflation.

In this section we apply our dynamical system formulation to give a rigorous analysis
of ${\cal R}$ and $\zeta$ in the present case of a minimally
coupled scalar field with exponential potential, and restricting
our attention to future accelerating models ($0\leq \lambda <1/\sqrt 3$, with
$\lambda=0$ giving a constant potential).
Earlier we showed that for these models as $N\rightarrow \pm \infty$
almost all orbits are asymptotic to fixed points in the long wavelength
boundary $Z=0$. The only exceptions are
the orbits that describe models with a constant deceleration parameter. These
orbits, which lie in the invariant sets $\Sigma_{\varphi}=\lambda$ and
$\Sigma_\varphi = \pm 1$, differ in that they are past/future asymptotic
to the short wavelength boundary ${\bar Z}=1$, respectively.

We proceed by using previous results
to derive asymptotic expressions for ${\cal R}$ and
$\zeta$ on approach to the fixed points labelled $\mathrm{M}_{\pm}$,
$\mathrm{A}_{\lambda}$, $\mathrm{S}_{\lambda}$,
i.e., as $N\rightarrow \pm \infty$ and $Z\rightarrow 0$.
In view of the heuristic conservation property of
${\cal R}$ and $\zeta$ one might conjecture that
the limits $\lim_{N\rightarrow \pm \infty} \cal R$
and $\lim_{N\rightarrow \pm \infty} \zeta$ would be finite and equal.
However, we have found that this does not hold in general,
since \emph{in two cases} ${\cal R}$ \emph{and} $\zeta$
\emph{become unbounded in the limit}.
In the first case, the conjecture fails for generic orbits
into the past for $\lambda$ satisfying $0\leq \lambda <1/\sqrt 3$,
and in the second case for generic orbits
into the future, but only for the ultra slow-roll case $\lambda=0$.
In both cases, however, there are non-generic orbits
for which the conjecture holds. In the first case we note that
generic orbits are past asymptotic to one of the fixed points
$\mathrm{M}_+$ or $\mathrm{M}_-$
and lie in the centre manifold of one of these fixed points.
However, the conjecture does hold for orbits that lie in the unstable manifold of
these fixed points, which is of one dimension less than the centre manifold.
In the second case we note that generic orbits with $\lambda=0$ are
asymptotic to the fixed point $\mathrm{A}_{0}$ into the future
while the conjecture does hold for a one-parameter family of orbits that is
asymptotic to $\mathrm{S}_{0}$.\footnote{The result of~\cite{rometal16} concerning
future non-conservation of ${\cal R}$ and $\zeta$
applies to the generic orbits with $\lambda=0$.}

To justify the above claims we derive asymptotic expressions for
${\cal R}$ and $\zeta$ as $N\rightarrow \pm \infty$.
We begin with the following relations, working in the total matter
(comoving) gauge:
\begin{equation} \label{Rzeta}
{\cal R} \equiv \psi_\mathrm{v},\qquad
\zeta\equiv\psi_{\rho}=\psi_\mathrm{v}-\frac13{\bdelta_\mathrm{v}},
\end{equation}
where the second equation is the change of gauge formula for the curvature
perturbation $\psi$.\footnote{See, for example,
Uggla and Wainwright (2019)~\cite{uggwai19a}, equation (49),
specialized to first order perturbations.}
For a perturbed scalar field one can show that\footnote{The
first equation is given as equation (2.11) in Uggla and
Wainwright (2019)~\cite{uggwai19c}. The second equation follows
from the perturbed Einstein equation $\psi_\mathrm{v}'=c_s^2{\bdelta_\mathrm{v}} +\Gamma$
(see equation (62a) in Uggla and Wainwright (2019)~\cite{uggwai18})
using the fact that $\Gamma=(1-c_s^2){\bdelta}_\mathrm{v}$
for a perturbed scalar field (see equation (3.3c) in Uggla and
Wainwright (2019)~\cite{uggwai19c}.)}
\begin{equation} \label{sf_relation}
\psi_\mathrm{v} = \frac{\varphi_\mathrm{c}}{\varphi_0'}, \qquad
\bdelta_\mathrm{v}=\psi_\mathrm{v}',
\end{equation}
We substitute~\eqref{sf_relation} into~\eqref{Rzeta} and use the dynamical
system variables $y=\varphi_\mathrm{c}'/\varphi_\mathrm{c}$ and $\varphi_0'=\sqrt 6\Sigma_{\varphi}$.
After rearranging we obtain\footnote{Note that
${\cal R}'/{\cal R}=\varphi_\mathrm{c}'/\varphi_\mathrm{c} - \Sigma_{\varphi}'/\Sigma_{\varphi}
=y - \Sigma_{\varphi}'/\Sigma_{\varphi}$.}
\begin{equation}\label{psidef}
{\cal R} = \frac{1}{\sqrt 6}\left(\frac{\varphi_\mathrm{c}}{\Sigma_{\varphi}}\right),\qquad
\zeta = {\cal R}\left(1-\frac13\left(y-\frac{\Sigma_{\varphi}'}{\Sigma_{\varphi}}\right)\right)
= {\cal R} - \frac13{\cal R}^\prime.
\end{equation}
We can now derive asymptotic expressions for ${\cal R}$ and $\zeta$
in the neighbourhood of the fixed points
for $\mathrm{M}_{\pm}$, $\mathrm{A}_{\lambda}$ and $\mathrm{S}_{\lambda}$
using the asymptotic expressions for $y$, $\varphi_\mathrm{c}$ and $\Sigma_{\varphi}$
given in the previous section.

We first consider orbits approaching the fixed points $\mathrm{M}_{\pm}$  into the past.
Keeping only the past dominant terms\footnote{The first terms omitted
have time dependence of the form $e^{2N}$ and $e^{4N}$.}
in the expansions in section~\ref{subsec:asympt} we obtain the relations:
\begin{subequations}
\begin{align}
\varphi_\mathrm{c} \approx C_{\varphi}^{-}(1+C_3 N), \quad
\Sigma_{\varphi}\approx \pm 1 \quad &\implies\quad {\cal R}\approx
\pm \frac{C_\varphi^-}{\sqrt{6}}(1 + C_3 N), \\
y-\frac{\Sigma_{\varphi}'}{\Sigma_{\varphi}} \approx \frac{C_3}{1+C_3 N} \quad
 &\implies \quad \zeta \approx  \pm\frac{C_\varphi^-}{\sqrt{6}}\left(1 + C_3 N - \frac13 C_3\right).
\end{align}
\end{subequations}
For generic orbits ($C_3\neq 0$) it follows that ${\cal R}$ and $\zeta$ diverge as
$N\rightarrow -\infty$, while for orbits on
the unstable manifold ($C_3 = 0$) we obtain
\begin{equation}
\lim_{N\rightarrow - \infty}{\cal R} = \lim_{N\rightarrow - \infty}\zeta
= \pm\frac{C_\varphi^-}{\sqrt{6}}.
\end{equation}

We next consider orbits approaching the fixed points $\mathrm{A}_{\lambda}$
and $\mathrm{S}_{\lambda}$ for $0 < \lambda <1/\sqrt 3$ into the future.
Keeping only the future dominant terms we obtain for $\mathrm{A}_{\lambda}$:
\begin{subequations}
\begin{align}
\varphi_c \approx C_{\varphi}^{+}, \quad
\Sigma_{\varphi} \approx \lambda \quad &\implies \quad \lim_{N\rightarrow\infty}{\cal R}=
\frac{C_{\varphi}^+}{\sqrt6 \lambda}, \\
y - \frac{\Sigma_{\varphi}'}{\Sigma_{\varphi}} \approx 0 \quad
&\implies \quad\lim_{N\rightarrow\infty}\zeta = \lim_{N\rightarrow\infty}{\cal R},
\end{align}
\end{subequations}
and for $\mathrm{S}_{\lambda}$:
\begin{subequations}
\begin{align}
\varphi_c \approx C_{\varphi}^{+} e^{-3(1-\lambda^2)N}, \quad
\Sigma_{\varphi}\approx \lambda \quad &\implies \quad  \lim_{N\rightarrow\infty}{\cal R}=0,\\
y-\frac{\Sigma_{\varphi}'}{\Sigma_{\varphi}} \approx -3(1-\lambda^2)\quad
&\implies \quad \lim_{N\rightarrow\infty}\zeta = (2-\lambda^2)\lim_{N\rightarrow\infty}{\cal R} =0.
\end{align}
\end{subequations}
This establishes that the limits of ${\cal R}$ and $\zeta$ are equal and finite in both cases.

We next consider orbits approaching the fixed points $\mathrm{A}_{\lambda}$ and
$\mathrm{S}_{\lambda}$ into the future, for the exceptional value $\lambda=0$.
Keeping only the future dominant terms we obtain for $\mathrm{S}_{0}$:
\begin{subequations}
\begin{align}
\varphi_c \approx C_{\varphi}^{+}e^{-3N},\quad
\Sigma_{\varphi}\approx C_1 e^{-3N} \quad &\implies\quad \lim_{N\rightarrow\infty}{\cal R}=
\frac{C_{\varphi}^+}{\sqrt6 C_1},\\
y-\frac{\Sigma_{\varphi}'}{\Sigma_{\varphi}} \approx 0 \quad
&\implies \quad \lim_{N\rightarrow\infty}\zeta =\lim_{N\rightarrow\infty}{\cal R},
\end{align}
\end{subequations}
and for $\mathrm{A}_0$:
\begin{subequations}
\begin{align}
\varphi_c \approx C_{\varphi}^{+},\quad
\Sigma_{\varphi}\approx C_1 e^{-3N} \quad &\implies\quad {\cal R} \approx
\frac{C_{\varphi}^+}{\sqrt6 C_1} e^{3N},\\
1-\frac{1}{3}\left(y-\frac{\Sigma_{\varphi}'}{\Sigma_{\varphi}}\right) \approx C_2 e^{-2N} \quad
&\implies \quad \zeta \approx \frac{C_{\varphi}^+ C_2}{3 \sqrt6 C_1} e^{N}.
\end{align}
\end{subequations}
This establishes that the limits of ${\cal R}$ and $\zeta$ are equal for the
fixed point $\mathrm{S}_0$, while for orbits that approach the fixed point
$\mathrm{A}_0$, ${\cal R}$ and $\zeta$ diverge as $N\rightarrow \infty$, but with
${\cal R}$ diverging faster than $\zeta$. In other words \emph {the conservation conjecture
breaks down for generic orbits toward the future}
since the fixed point $\mathrm{A}_0$ is the future attractor, but
it is valid for the restricted set of orbits that approach $\mathrm{S}_0$.

The latter result is related to those in \cite{rometal16} where it
is shown that  conservation of $\cal R$ and $\zeta$  for long wavelength
perturbations breaks down in the case of ultra slow-roll inflation toward
the future (see also~\cite{moopal15,tsawoo04,kin05,nametal13,cheetal13a,maretal13}).
We can link our work to theirs by keeping additional terms in the expansion for
$\varphi_c$ and $y$.\footnote{This yields
$\varphi_c\approx C_{\varphi}^+(1-\sfrac13e^{-3N} +\sfrac12 e^{-2N}),\,
y\approx -C_2e^{-2N} +C_3e^{-3N}$.}
This leads to
\begin{subequations}
\begin{align}
{\cal R} &\approx \frac{C_\varphi^+}{3\sqrt{6}\,C_1}
\left[3e^{3N}\left(1 + \frac{C_2}{2}e^{-2N}\right)
 - C_3\right],\\
\zeta &\approx \frac{C_\varphi^+}
{3\sqrt{6}\,C_1}\left[e^{N}C_2 - C_3\right].
\end{align}
\end{subequations}
These expansions are equivalent to equations (36) and (37) in~\cite{rometal16},\footnote{Note
that $a^3=e^{3N}$ and $k^2/{\cal H}^2\equiv Z\approx C_2e^{-2N}$ as
$N\rightarrow \infty$.} which in~\cite{rometal16} were derived using the usual heuristic approach to
long wavelength perturbations.\footnote{The present dynamical systems approach can be
used to obtain more refined approximations by making Picard expansions
(see e.g.~\cite{ugg89} and references therein), i.e., expansions based on the
eigenvalues associated with the fixed points in the dynamical system.
In, e.g., the case $\lambda = 0$ and $\mathrm{A}_0$ this entails an expansion based on
$e^{-2N}$ and $e^{-3N}$ where the equations for $(\Sigma_\varphi, Z, y)$
are solved for each coefficient of $e^{-2N}, e^{-3N}, e^{-4N}, e^{-5N}, e^{-6N},...$,
which when inserted into~\eqref{psidef} leads to expansions in $C_1$,
$C_2 \propto k^2$, $C_3$ and since only $C_2$ involves $k^2$ correct approximate
solutions do not involve series expansions in $k^2$ alone.}

\begin{figure}[ht!]
     \begin{center}
         \subfigure[State-space]{\label{} \includegraphics[width=0.44\textwidth]{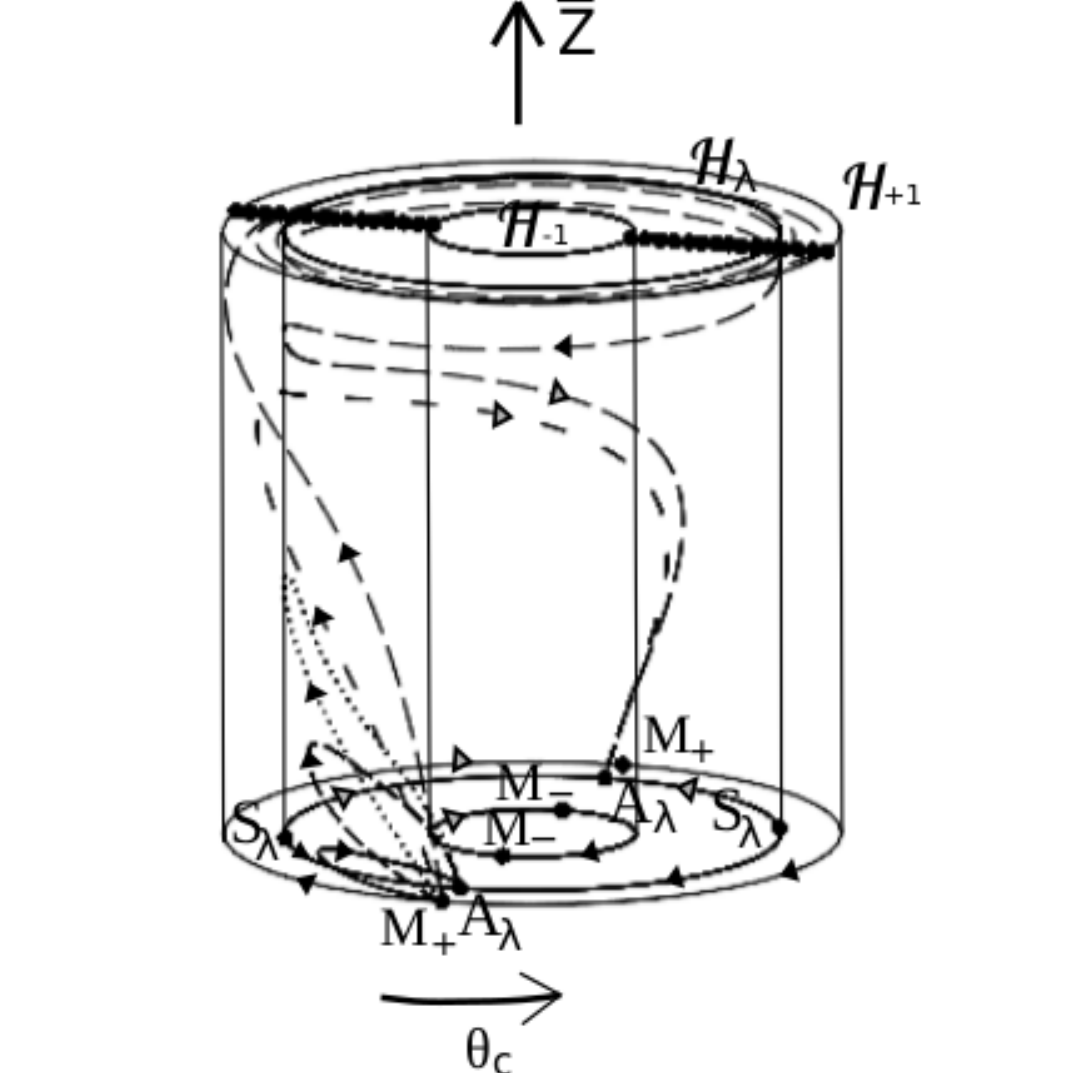}}\hspace{0.8cm}
         \subfigure[$\bar{Z}(N)$]{\label{}
             \includegraphics[width=0.45\textwidth]{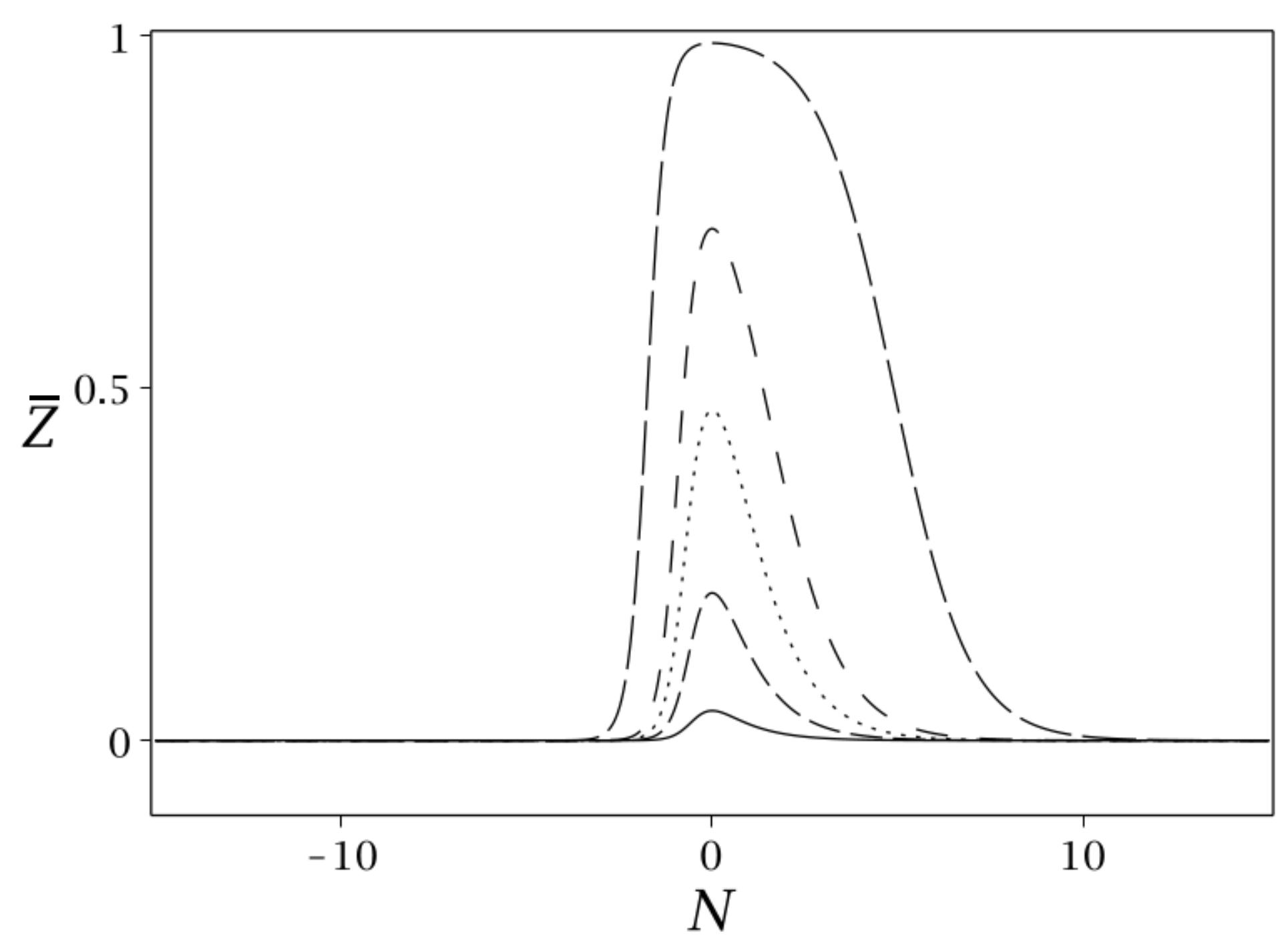}}
          \\
         \subfigure[$\mathcal{R}(N)$]{\label{}
             \includegraphics[width=0.44\textwidth]{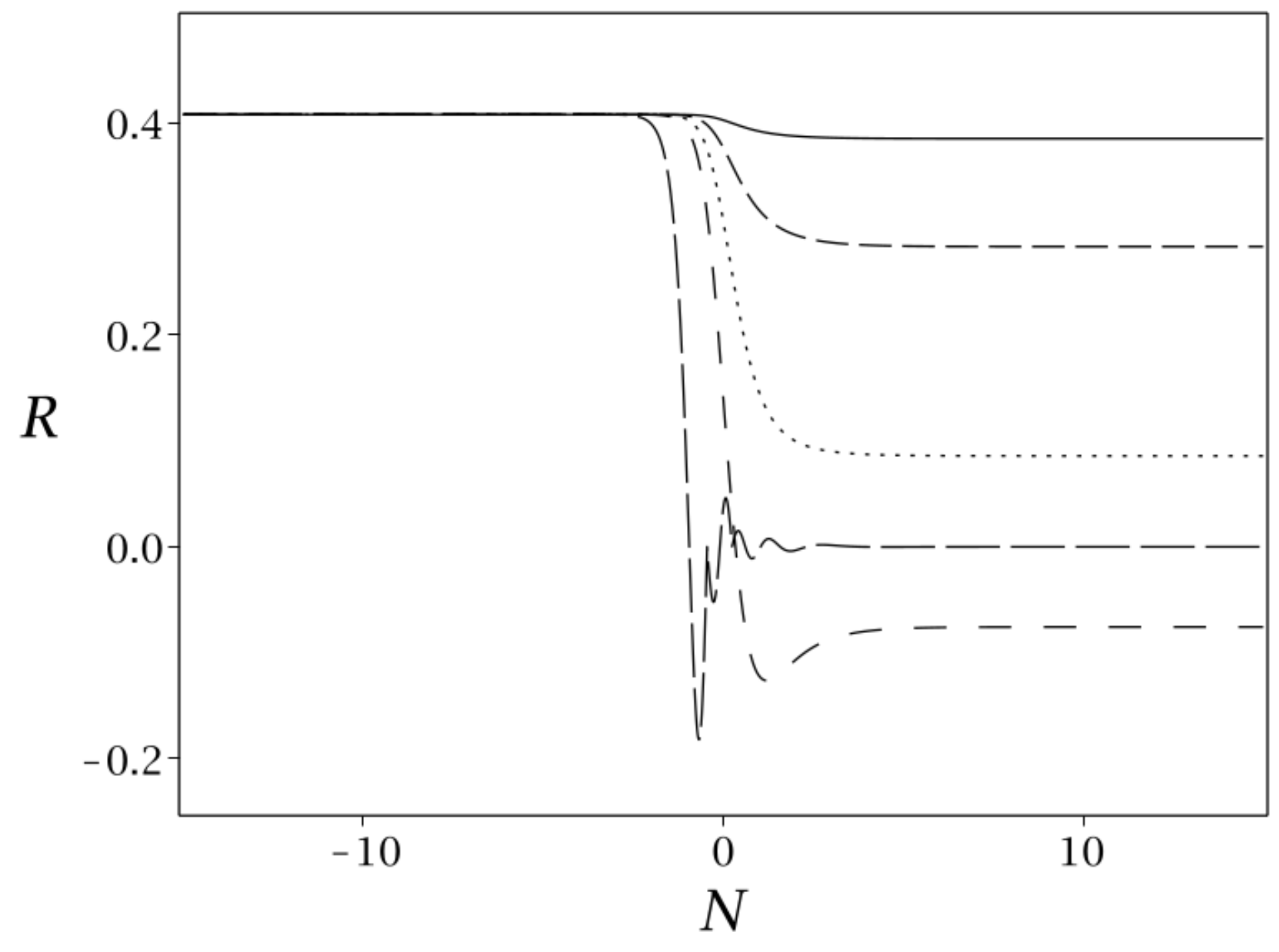}}\hspace{1cm}
         \subfigure[$\zeta(N)$]{\label{} \includegraphics[width=0.45\textwidth]{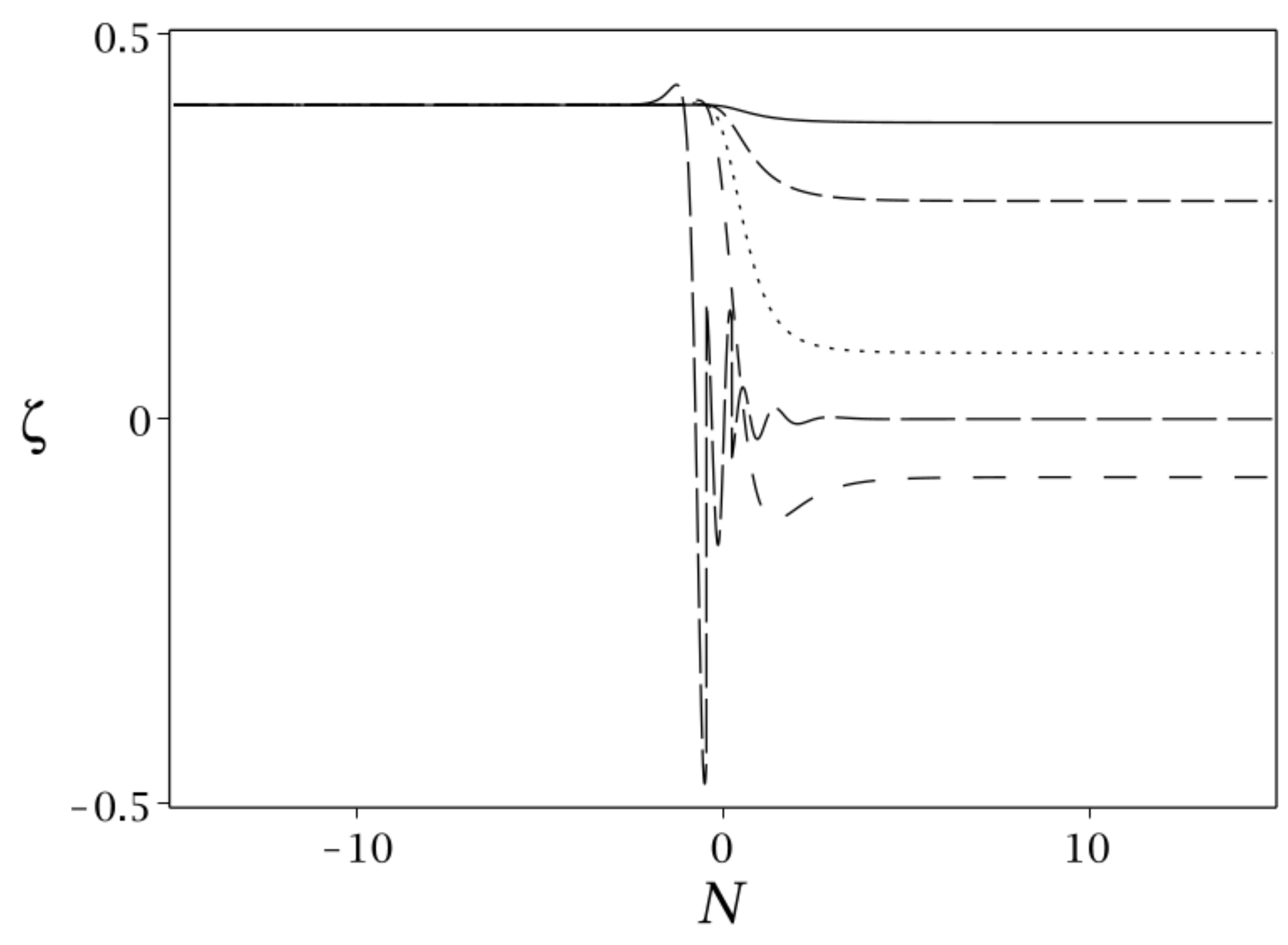}}
         \vspace{-0.5cm}
     \end{center}
     \caption{Orbits on the unstable manifold of $\mathrm{M}_+$ when $\lambda=1/\sqrt{6}$, and the associated graphs
     for $\bar{Z}$, ${\cal R}$ and $\zeta$.
}\label{FigCurvPert}
\end{figure}

Let us conclude this section by showing some of the global properties of ${\cal R}$ and
$\zeta$ by numerically plotting some orbits in the state space and then illustrating
the properties these solutions give rise to for ${\cal R}$ and $\zeta$.
Since the ultra slow-roll case $\lambda=0$ yields generic unboundedness both toward the
past and future, we restrict considerations to the case $0< \lambda < 1/\sqrt{3}$,
which is represented by $\lambda = 1/\sqrt{6}$. We then note that ${\cal R}$ and $\zeta$
become unbounded generically toward the past; it is only the unstable manifold of $\mathrm{M}_\pm$
that yield past boundedness of these quantities. For this reason we restrict our considerations
to the orbits on the unstable manifold of $\mathrm{M}_+$, see Fig.~\ref{FigCurvPert}.
Note that for sufficiently small $Z_\mathrm{max}$, ${\cal R}$ and $\zeta$ are globally
approximately conserved in this case.
%

\section{Discussion\label{sec:disc}}

We now indicate how our dynamical systems approach can be
extended to apply to non-exponential scalar field potentials for which
$\lambda(\varphi)$, defined by~\eqref{defback1},
is bounded.\footnote{For how to globally deal with backgrounds with unbounded $\lambda(\varphi)$,
see~\cite{alhetal15,alhugg15,alhugg17}.} Equation~\eqref{backgroundeq} describes the evolution
of the background scalar field $\varphi$  through the first order differential
equations for $\varphi$ and $\Sigma_\varphi$. For an exponential potential,
we have seen that the equation for $\varphi$
decouples, but for a general potential one has to keep both $\varphi$ and $\Sigma_\varphi$ as variables in the background
state space $\cal B$.
In order to obtain a compact background state space
we introduce a new bounded scalar field variable
\begin{equation} \label{bar.varphi}
\bar{\varphi} = f(\varphi),\qquad -\infty<\varphi<\infty,
\end{equation}
where $f$ is a monotone increasing, differentiable and bounded function.
We define $\bar{\varphi}_\pm = \lim_{{\varphi}\rightarrow\pm\infty}f(\varphi)$,
and also require that
\begin{equation}
\lim_{\varphi\rightarrow \pm \infty}\left(\frac{df}{d\varphi}\right)=0.
\end{equation}
This results in coupled differential equations of the form
\begin{subequations}\label{backgrounde2}
\begin{align}
\bar{\varphi}^\prime &= \sqrt{6}\left(\frac{d\bar{\varphi}}{d\varphi}\right)\Sigma_\varphi ,\label{barphisig}\\
\Sigma_\varphi^\prime &= -3(1 - \Sigma_\varphi^2)(\Sigma_\varphi - \lambda),\label{Sigmaeq2}
\end{align}
\end{subequations}
where $\lambda$ and $d\bar{\varphi}/d\varphi$ are expressed
as (differentiable) functions of $\bar{\varphi}$ using the
inverse $\varphi =f^{-1}(\bar{\varphi})$.

Thus the main change in generalizing from an exponential potential
to a potential $V(\varphi)$ with bounded $\lambda(\varphi)$ is that a single
differential equation for $\Sigma_{\varphi}$ is replaced by two coupled
differential equations~\eqref{backgrounde2} for $\bar{\varphi}$ and $\Sigma_{\varphi}$,
which means that \emph{the dimension of the background state space $\cal B$
increases from two to three} (the third variable is ${\bar Z}$).
The dimension of the perturbation state space ${\cal P}$  is unchanged
at one: initially there are two perturbation variables but after compactification
${\cal P}$ is identified with a circle $S^1$. The perturbed KG equation,
given by~\eqref{KG.general}, has the same general structure,
but the coefficients are now determined by the background quantities,
$Z=k^2{\cal H}^{-2}$, $\Sigma_\varphi$,
$\lambda(\varphi)$, and a new function $\Upsilon(\varphi)$, defined by
\begin{equation}
\Upsilon(\varphi) = \frac{V_{,\varphi\varphi}}{6V},
\end{equation}
which we also assume is bounded. In the case of an
exponential potential $\Upsilon = \lambda^2$, and
hence is constant.

If the scalars $\lambda$ and $\Upsilon$ are not constant but satisfy
\begin{equation}  \label{asymp.exp}
\lim_{\varphi\rightarrow \pm\infty}\lambda = \lambda_\pm, \qquad
\lim_{\varphi\rightarrow \pm\infty}\Upsilon = \lambda_\pm^2,
\end{equation}
where $\lambda_{\pm} = \mathrm{const}.$ we say that
\emph{the potential $V(\varphi)$ is asymptotically exponential in the limit
$\varphi\rightarrow \pm\infty$} and hence as
$\bar{\varphi} \rightarrow \bar{\varphi}_\pm$.
For a such a potential the background state
space ${\cal B}$ is defined by the inequalities $-1\leq \Sigma_{\varphi} \leq 1, \,
{\bar \varphi_{-}}\leq  {\bar \varphi}\leq {\bar \varphi_{+}},\, 0\leq {\bar Z}\leq 1$.
The boundary has six components given by
$\Sigma_{\varphi} = \pm 1, \, {\bar \varphi} =  {\bar \varphi_\pm} ,\,  {\bar Z}= 0, 1$,
that are invariant sets.
On account of~\eqref{asymp.exp} the dynamical system
that is defined on the boundary components
${\bar \varphi} = {\bar \varphi_\pm}$ coincides with the dynamical
system that governs a perturbed scalar field with exponential potential,
with the parameter $\lambda$ given by $\lambda=\lambda_{\pm}$ .
In this way the dynamical system developed in this
paper for a scalar field with an
exponential potential acts as a building block for dynamical
systems that describe a scalar field with asymptotically
exponential (or constant) potentials.

The choice of the function $f(\varphi)$ in equation~\eqref{bar.varphi} depends
on the form of the potential, and finding
a suitable function involves some experimentation.
We illustrate the process by considering a scalar field model
introduced by Dimopoulos and Owen (2017) in~\cite{dimowe17},
where the  potential is given by
\begin{equation} \label{dimowe.pot}
V(\varphi) = V_*\left\{e^{\kappa\beta\left(1-\tanh{\frac{\varphi}{\beta}}\right)}-1\right\},
\end{equation}
depending on the constants $V_*$, $\kappa$ and $\beta$, with $\beta>0$ and
$e^{ 2\kappa\beta}>1$.\footnote{$\beta = \sqrt{6\alpha}$ in the
notation used in~\cite{dimowe17}.}
This model describes so-called quintessential inflation,\footnote{See
Peebles and Vilenkin (1999)~\cite{peevil99}.}
in which the scalar field
creates two phases of accelerated expansion, one at early times which leads to
inflation and the other at late times which leads to quintessence.
This potential is asymptotically exponential (even constant) since it can be verified that
\begin{equation} \label{infl.plateau}
V(\varphi) \approx(e^{ 2\kappa\beta}-1)V_*, \qquad
 \varphi\rightarrow -\infty,
\end{equation}
which approximates an exponential potential with $\lambda_-=0$, and
\begin{equation} \label{quin.plateau}
V(\varphi) \approx 2\kappa\beta V_*e^{-2\varphi/\beta},\qquad
 \varphi\rightarrow \infty, \quad
\end{equation}
which approximates an exponential potential with $\lambda_+=\sqrt{\sfrac23}\beta^{-1}$.
The potential thus has an inflationary
plateau described by~\eqref{infl.plateau} and a quintessential tail described
by~\eqref{quin.plateau}, see figure 1 in~\cite{dimowe17}.

For this potential it is convenient to define the function $f(\varphi)$
in equation~\eqref{bar.varphi} by
\begin{equation}
\bar{\varphi} = \tanh{\frac{\varphi}{\beta}},
\end{equation}
so that $\bar{\varphi}_\pm = \lim_{{\varphi}\rightarrow\pm\infty}\bar{\varphi} = \pm 1$.
It follows that
\begin{equation}
\frac{d\bar{\varphi}}{d \varphi} = \beta^{-1}(1-\bar{\varphi}^2),
\end{equation}
as required in~\eqref{barphisig}, and one can verify that $\lambda(\bar{\varphi})$ and
$\Upsilon(\bar{\varphi})$ are bounded and differentiable on $\bar{\varphi} \in [-1,1]$.
Moreover, it can be verified that  $\lambda(\bar{\varphi})$ and
$\Upsilon(\bar{\varphi})$ satisfy~\eqref{asymp.exp}, with
$\lambda_+ =\sqrt{\sfrac23}\beta^{-1}$ and $\lambda_- =0,$ which confirms
that the potential~\eqref{dimowe.pot} is asymptotically exponential (constant) when
$\varphi \rightarrow +\infty$ ($\varphi \rightarrow -\infty$). The perturbation space
for this potential will be treated in a forthcoming paper.

We conclude with some brief remarks on the structure of
the state space $\cal S$=${\cal B}\times {\cal P}$  for models with multiple sources.
In a future paper we will generalize the analysis of a scalar field with
exponential potential in this paper
by adding dust (CDM) as a second source, with the
two sources assumed to be non-interacting. The main change will be to add
the density parameter for the dust, $\Omega_m$, as a second background matter variable.
The background state space ${\cal B}$ will thus be three dimensional, with
coordinates $(\Sigma_{\varphi}, \Omega_m, \bar{Z})$, with $\bar{Z}$
describing the evolution of the background geometry represented by
${\cal H}$, as in the present paper.

There is, however,  a significant
increase in complexity as regards the perturbation space ${\cal P}$.
In the present paper, one starts with two perturbation variables,\footnote
{Initially there is in fact one perturbation variable $f$ that satisfies a second order
differential equation. In order to derive a dynamical system we consider $f,f'$
as two independent perturbation variables, which then
define one angular variable $\theta$ using $f'=f \tan\theta.$ }
but the compactified state space is the circle $S^1$.
With two sources there will initially be four scalar perturbation variables
(two for each component) and the process of compactification will lead to a
three-sphere $S^3$ that is defined by a constraint equation. The perturbation
space ${\cal P}$ will thus be three dimensional and compact, or four
dimensional with one constraint.

\subsection*{Acknowledgments}
AA is supported by CAMGSD, Instituto Superior T{\'e}cnico by FCT/Portugal through UID/MAT/04459/2019 and UIDB/MAT/04459/2020 and project (GPSEinstein)  PTDC/MAT-ANA/1275/2014. CU would like to thank the CAMGSD, Instituto Superior T{\'e}cnico in Lisbon and the University of Waterloo, Canada, for kind hospitality.

\bibliographystyle{unsrt}
\bibliography{../Bibtex/cos_pert_papers}

\end{document}